\documentclass[fleqn,usenatbib]{mnras}
\usepackage[T1]{fontenc}

\usepackage{graphicx}	
\usepackage{amsmath}	
\usepackage{amssymb}	
\usepackage{comment}
\usepackage{bbold}

\usepackage{pifont}%
\newcommand{\cmark}{{\color{green}\ding{51}}}%
\newcommand{\xmark}{{\color{red}\ding{55}}}%

\usepackage{bm}
\usepackage{subfigure}
\usepackage{tabularx}
\usepackage[english=american]{csquotes}
\usepackage{hyperref}
\hypersetup{
    colorlinks=true,
    linkcolor=red,
    citecolor=blue,
    urlcolor=blue,
}

\newcommand{\cmfast}{\textsc{{\tt 21cmFAST}}}

\title[21-cm likelihood]{Exploring the likelihood of the 21-cm power spectrum with simulation-based inference}

\author[Prelogović \& Mesinger]{
David Prelogović\thanks{E-mail: david.prelogovic@sns.it} \&
Andrei Mesinger
\\
Scuola Normale Superiore, Piazza dei Cavalieri 7, 56125 Pisa, Italy\\
}

\date{Accepted XXX. Received YYY; in original form ZZZ}

\pubyear{2021}

\begin{document}
\label{firstpage}
\pagerange{\pageref{firstpage}--\pageref{lastpage}}
\maketitle

\begin{abstract}
Observations of the cosmic 21-cm power spectrum (PS) are starting to enable precision Bayesian inference of galaxy properties and physical cosmology, during the first billion years of our Universe.
Here we investigate the impact of common approximations about the likelihood used in such inferences, including: (i) assuming a Gaussian functional form; (ii) estimating the mean from a single realization; and (iii) estimating the (co)variance at a single point in parameter space.
We compare \enquote{classical} inference that uses an explicit likelihood  with simulation based inference (SBI) that estimates the likelihood from a training set.
Our forward-models include: (i) realizations of the cosmic 21-cm signal computed with \cmfast{} by varying UV and X-ray galaxy parameters together with the initial conditions; (ii) realizations of the telescope noise corresponding to a $1000 \, \mathrm{h}$ integration with SKA1-Low; (iii) the excision of Fourier modes corresponding to a foreground-dominated, horizon \enquote{wedge}. 
We find that the 1D PS likelihood is well described by a Gaussian accounting for covariances between wavemodes and redshift bins (higher order correlations are small).  However, common approaches of estimating the forward-modeled mean and (co)variance from a random realization or at a single point in parameter space result in biased and over-constrained posteriors.  
Our best results come from using SBI to fit a non-Gaussian likelihood with a Gaussian mixture neural density estimator.
Such SBI can be performed with up to an order of magnitude fewer simulations than classical, explicit likelihood inference.  Thus SBI provides accurate posteriors at a comparably low computational cost.
\end{abstract}




\section{Introduction}
The redshifted 21-cm signal, corresponding to the hyperfine splitting of the 1S ground state of neutral hydrogen, is becoming a powerful probe of the first billion years of our Universe.  This evolutionary milestone includes the cosmic dawn (CD) of the first galaxies and the eventual reionization of the intergalactic medium (IGM).   Containing orders of magnitude more independent modes than the CMB (e.g. \citealt{Loeb_2004}), this cosmic 21-cm signal will allow us to infer the properties of the unseen first galaxies (e.g. \citealt{Park_2019}), together with physical cosmology (e.g. \citealt{Kern_2017}). Current radio telescopes are already putting compelling upper limits on its spherically-averaged power spectrum (PS) (e.g. \citealt{Trott_2020, Mertens_2020, HERA_2022b, HERA_2022a}). The upcoming Square Kilometer Array (SKA)\footnote{\url{https://www.skatelescope.org}} should not only \textit{detect} the PS, but also enable image-space mapping of the signal \citep{Koopmans2015, Mesinger2020}.

There are several important reasons why these experiments primarily aim to detect the PS (as compared with other summary statistics). Firstly, it is the most physically motivated summary: the theoretical PS can be analytically computed for simple functions of the matter field which is Gaussian on large-scales (e.g. \citealt{Barkana2005, Pritchard2007, McQuinn2018, Schneider2021, Munoz2023}).
Secondly, due to its high compression (many Fourier modes are sampled to compute the PS in a given bin), it provides a good signal-to-noise ratio (S/N) facilitating a detection with limited, preliminary data. Finally, the visibility space of interferometeric measurements is closely related to Fourier space.  This means that the telescope noise, systematics, and foreground contamination can be more naturally characterized in PS space, compared with e.g. image based summaries.  As a result, inference from the 21-cm signal generally uses the PS when computing a likelihood (though see, e.g. \citealt{shimabukuro2017constraining, Gazagnes_2021, Watkinson_2022, Greig_2023}).

What is the form of the likelihood of the 21-cm PS?
Virtually all works (e.g. \citealt{Greig_2015, Ghara_2020, Greig_2021, Maity_2023, HERA_2023}) assume a Gaussian form for the likelihood:
\begin{align}
\label{eq:gaussian_likelihood}
\ln \mathcal{L}(\Delta^2_{21 \, \text{obs}} | \bm{\theta}) &\propto
-\frac{1}{2} \, [\Delta^2_{21 \, \text{obs}} - \mu(\bm{\theta})]^T \, \Sigma^{-1} \, [\Delta^2_{21 \, \text{obs}} - \mu(\bm{\theta})] \nonumber \\
&\approx -\frac{1}{2} \sum_{k, z}\frac{[\Delta^2_{21 \, \text{obs}}(k, z) - \mu(\bm{\theta}, k, z)]^2}{  \sigma^{2}(\bm{\theta}, k, z)} ,
\end{align}
where $\Delta^2_{21 \, \text{obs}}(k, z)$ is the \enquote{observed} PS (c.f.  Eq.~\ref{eq:1DPS}).  For a given astrophysical/cosmological parameter vector $\bm{\theta}$, the forward-modeled PS is {\it stochastic}, primarily due to cosmic variance and thermal noise: $\mu(\bm{\theta}, k, z)$ is the  mean of this  distribution and $\Sigma(\bm{\theta}, {k, z, k', z'}$) its covariance matrix capturing 2-pt correlations between wavemodes and redshifts (note we omit the explicit summation over bins, writing it as a vector-matrix product for simplicity).
Most often, the full covariance matrix is approximated to be diagonal with variance $\sigma^2(\bm{\theta}, k, z)$ (though see \citealt{Nasirudin_2020}), leading to the simple summation over $(k, z)$ bins.  Furthermore, due to computational requirements, the forward-modeled mean and (co)variance are not computed on-the-fly at each $\bm{\theta}$ by varying the cosmic seed and other sources of stochasticity.  Instead, a single realization is used in place of the mean, and the (co)variance is generally only computed at a single, fiducial parameter, $\bm{\theta}_{\rm fid}$, and then kept fixed during inference (though see \citealt{Watkinson_2022}). 
One then infers the posterior probability distribution $P(\bm{\theta} | \Delta^2_{21 \, \text{obs}})$, using the Bayes' equation:
\begin{equation}
    P(\bm{\theta} | \Delta^2_{21 \,  \text{obs}}) = \frac{\mathcal{L}(\Delta^2_{21 \,  \text{obs}} | \bm{\theta}) \cdot \pi(\bm{\theta})}{\mathcal{Z}(\Delta^2_{21 \,  \text{obs}})} \, ,
    \label{eq:bayes_eq}
\end{equation}
where $\pi(\bm{\theta})$ the prior probability of $\bm{\theta}$, and the evidence $\mathcal{Z}(\Delta^2_{21 \, \text{obs}})$ is here treated as a normalization constant as we are focusing only on a single model.

The form of the likelihood in Eq. (\ref{eq:gaussian_likelihood}) is only guaranteed to be true for a Gaussian random field.  Although the primordial density fluctuations are indeed Gaussian, subsequent non-linear evolution through gravitational collapse and radiative cooling introduces non-Gaussianities in the hydrogen density field on moderate to small scales.  More importantly, the cosmic 21-cm signal is also sensitive to the temperature and ionization state of the IGM, which is in turn driven by the multi-wavelength radiation fields of the first galaxies.  Indeed, this dependence is precisely what allows us to infer the properties of these galaxies; however, it introduces highly non-trivial mode coupling across a wide range of scales.  As a result, the redshifted 21-cm signal from the CD is {\it not} Gaussian (e.g. \citealt{Mellema2013, Watkinson_2022, Greig_2022}).  This has two important implications.  Firstly, the PS does not provide optimal compression; other summary statistics could in principle better constrain astrophysical and cosmological parameters (e.g. \citealt{Gazagnes_2021, Zhao2022, Greig_2023}; Prelogović \& Mesinger, in prep).  Secondly, the likelihood of the PS is {\it not} Gaussian distributed according to Eq.~(\ref{eq:gaussian_likelihood}).  Quantifying this later point is the subject of this work.

How can one determine the form of the 21-cm PS likelihood?  Ideally, one would start with the Gaussian initial conditions and trace the likelihood throughout the evolution of the Universe, either analytically or numerically. Indeed, the field of large scale structure (LSS) as probed by galaxy surveys has made impressive progress along those lines  
(e.g. \citealt{Kitaura2008, Jasche_2010, Jasche_2012, Jasche_2013, Leclercq_2017, McAlpine_2022, Dai_2022, Bayer2022}). 
However, the dependence on unknown, multi-scale, multi-wavelength radiation fields makes it more difficult to obtain a tractable likelihood for the 21-cm PS.

Another option is to use simulation based inference (SBI; for a recent review see \citealt{cranmer_2020}). SBI relies on a training set of simulated data, which is used to train neural density estimators (NDE) to fit the likelihood without having to specify its functional form (indeed SBI is commonly also referred to as \enquote{likelihood-free} or \enquote{implicit likelihood} inference).
Specifically, SBI consists of the following general steps:
\begin{itemize}
    \item Sample a parameter vector from the prior $\widetilde{\bm{\theta}} \sim \pi(\bm{\theta})$.
    \item Simulate a data vector corresponding to the parameter sample, $\widetilde{\bm{d}} = \operatorname{simulator}(\widetilde{\bm{\theta}})$. This is equivalent to drawing from the likelihood function $\widetilde{\bm{d}} \sim \mathcal{L}(\bm{d} | \widetilde{\bm{\theta}})$.
    \item Repeat many times.
\end{itemize}
Here the data space $\bm{d}$ to corresponds to the 21-cm power spectrum $\Delta^2_{21} (k, z)$. For the simulator of the 21-cm signal, here we use the public {\tt 21cmFAST} code, described in more detail in section \ref{ch:21cm_inference}. The resulting set 
$\left\{(\widetilde{\bm{d}}_1, \widetilde{\bm{\theta}}_1), \ldots, (\widetilde{\bm{d}}_N, \widetilde{\bm{\theta}}_N)\right\}$ corresponds to samples from the joint distribution $P(\bm{d}, \bm{\theta}) = \mathcal{L}(\bm{d} | \bm{\theta}) \cdot \pi(\bm{\theta})$. One can then use a density estimator to fit the likelihood $\mathcal{L}(\bm{d} | \bm{\theta})$. 
Note that there are many variants to such a procedure (e.g. \citealt{cranmer_2020}), including fitting the posterior directly or fitting likelihood ratios (e.g. \citealt{Cole2022}).
SBI is rapidly becoming popular in the field of cosmic 21-cm due to its ability to deliver efficient, unbiased inference \citep{Zhao2022, Zhao2022b, Saxena2023}.

In this work, we use SBI to test the validity of a wide range of approximations for the likelihood of the spherically-averaged 21-cm PS.  These include: (i) a 1D variance evaluated at a single parameter choice; (ii) a 2D covariance evaluated at a single parameter choice; (iii) a 1D variance evaluated on-the-fly; (iv) a 2D covariance evaluated on-the-fly; (v) a fully non-Gaussian likelihood (allowing us to capture all higher order moments between $k$ and $z$ bins).  We compare approaches on the basis of both accuracy (verified in some cases using simulation based calibration; SBC), as well as computational efficiency.  As this work was nearing completion, a similar study was published by \cite{Zhao2022b}, comparing SBI to explicit likelihood inference in the recovery of two astrophysical parameters from the 21-cm PS. Our work is an improvement because: (i) we use a larger astrophysical parameter space; (ii) we compute all summaries on the lightcone (instead of computing some on lightcones and others on coeval cubes), thus allowing for a more precise comparison between different likelihood estimators; (iii) we compute the noise more accurately based on baseline sampling directly in $uv$ space; (iv) we explore a broader set of functional forms for the likelihood. 

This paper is organized as follows. In section \ref{ch:21cm_inference} we discuss the main ingredients of the inference, including our forward simulator, ${\tt 21cmFAST}$, astrophysical parameters and priors. In section \ref{ch:choosing_a_likelihood} we discuss the various functional forms of the PS likelihood we explore as well as the neural density estimators (NDEs) used to fit them. In \S\ref{ch:results} we show results for a fiducial mock observation, and test the inference across our parameter space with simulation based calibration. Finally, we conclude in \S\ref{ch:conclusions}. All quantities are quoted in co-moving units, and we assume a standard $\Lambda\mathrm{CDM}$ cosmology: $\left(\Omega_{\Lambda}, \Omega_{\mathrm{M}}, \Omega_{b}, n, \sigma_{8}, H_{0}\right) =\left(0.69,0.31,0.048,0.97,0.81,68 \mathrm{~km} \mathrm{~s}^{-1} \mathrm{Mpc}^{-1}\right)$, consistent with the results from \cite{Planck_2018VI}. 

\section{Simulating 21-cm observations} \label{ch:21cm_inference}

\begin{figure}
    \centering
    \includegraphics[width=.99\linewidth]{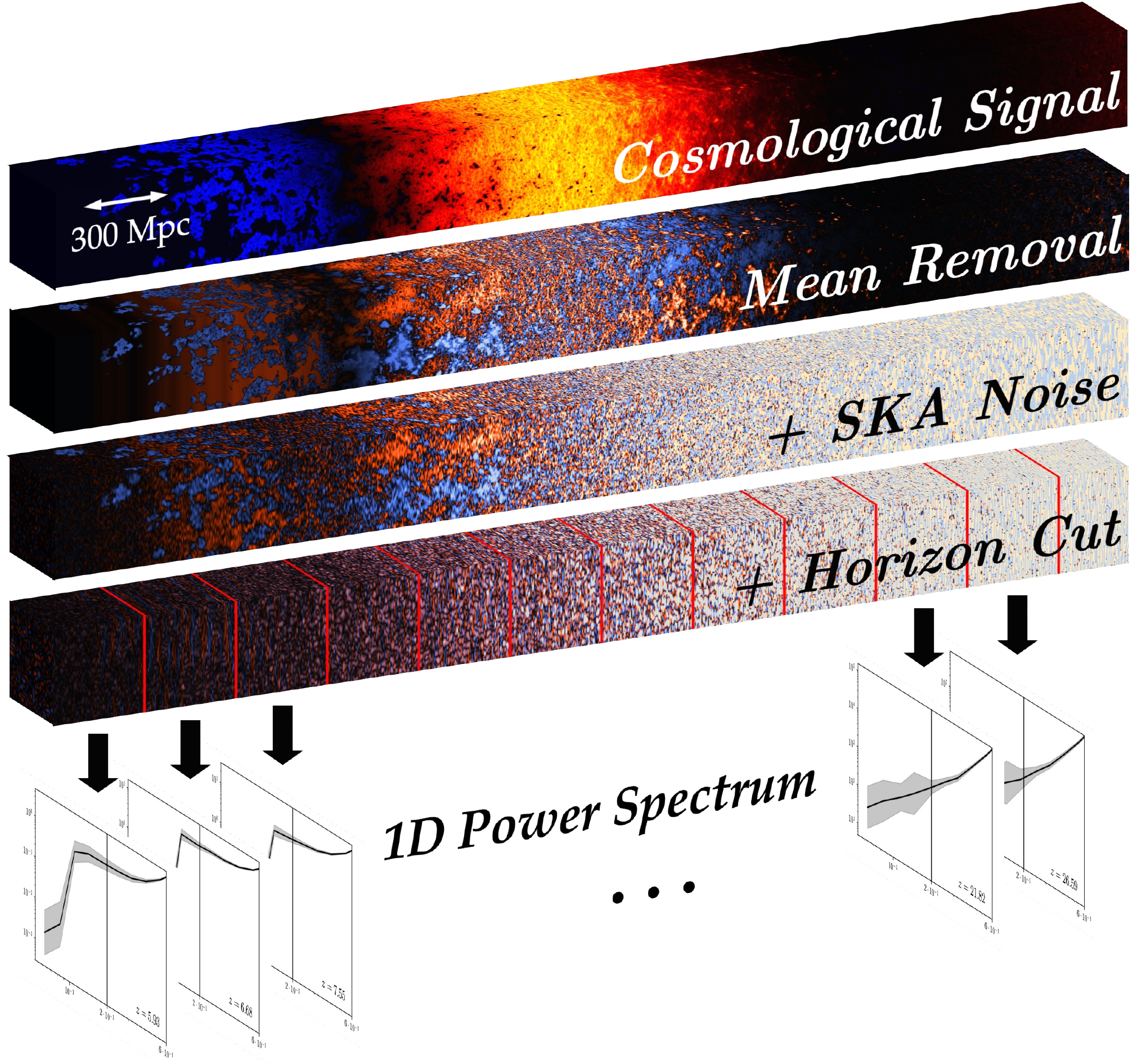}
    \caption{Schematic of our simulation pipeline. Starting from the cosmological signal computed with \cmfast{}, we remove the mean of the signal, add noise corresponding to a $1000 \mathrm{h}$ SKA1-Low observation, and perform a foreground cut below the horizon limit (see text for details).  Finally, we bin the lightcone and compute the 1D power spectrum in each bin. For the reference, in the upper left corner of the first lightcone, we show the scale of the simulation box size.}
    \label{fig:pipeline}
\end{figure}

Inference, whether using SBI or with an explicit likelihood, requires an accurate simulator to generate mock observables from samples of astrophysical/cosmological parameters.  Our simulation pipeline is illustrated in Figure \ref{fig:pipeline}. It consists of the following steps:
\begin{itemize}
    \item \textit{Cosmological signal} - we simulate a  realization of the 21-cm lightcone (more precisely a light "cuboid"), corresponding to a sampled parameter vector, $\tilde{\theta}$, and a sampled random seed for generating the initial conditions.
    \item \textit{Mean removal} - we remove the mean from each redshift/frequency slice, to account for the inability of interferometers to measure the $k_\perp=0$ mode.
    \item \textit{+ SKA Noise} - we add a realization of noise corresponding to a $1000 \mathrm{h}$ integration with SKA1-Low.
    \item \textit{+ Horizon cut} - we remove a foreground-dominated \enquote{wedge} region by zeroing the corresponding Fourier modes.
    \item \textit{1D Power Spectrum} - we cut the lightcone  into blocks of equal conformal length along the redshift axis, computing the 1D PS for each block.  This results in $\Delta^2_{21}(k, z)$ that we use as our summary statistics throughout this paper.
\end{itemize}
Below we briefly describe these steps in more detail.

To compute the cosmological signal from the first step, we use the public, semi-numerical code \texttt{\cmfast{} v3}\footnote{\url{https://github.com/21cmfast/21cmFAST}} \citep{21cmFAST_Mesinger11, 21cmFAST_JOSS}, with the galaxy parametrization from \cite{Park_2019}. 
The code generates a realization of the initial density and velocity fields, and evolves them with second order Lagrangian perturbation theory (2LPT; \citealt{Scoccimarro_1998}). From the evolved density fields, the conditional halo mass function is used to compute the spatial fluctuations in the galaxy field (e.g. \citealt{barkana2004}). The inhomogeneous reionization field is obtained by comparing the number of ionizing photons to the number of recombinations, in regions of decreasing radii \citep{furlanetto2004, sobacchi2014}. Soft UV and X-ray photons that have much longer mean free paths are instead tracked by integrating the local emissivity back along the lightcone, for each simulation cell.  These radiation fields then impact the temperature and ionization state of each IGM cell.  For more details, interested readers are encouraged to see \citep{21cmFAST_Mesinger_07, 21cmFAST_Mesinger11}.\textbf{}  

Our simulations correspond to 300  Mpc boxes, with a cell size of 1.5 Mpc. The choices of astrophysical galaxy parameters are discussed in the following section.  We interpolate between adjacent comoving snapshots, also accounting for sub-grid redshift space distortions (e.g. \citealt{Mao_2012, Jensen2013, Greig_2018}), creating a lightcone of the cosmic signal extending from redshift 30 to 5 (see the top panel of Fig. \ref{fig:pipeline}).

SKA1-Low $uv$ coverage and thermal noise are calculated using \texttt{tools21cm}\footnote{\url{https://github.com/sambit-giri/tools21cm}} \citep{tools21cm_JOSS}. We assume a tracked scan of $6 \mathrm{h}$ per day, $10 \mathrm{s}$ integration time, for a total of $1000 \mathrm{h}$, using only the core stations (baseline $\le 2 \mathrm{km}$). After subtracting the mean signal from each slice in the lighcone and adding the thermal noise corresponding to this $uv$ coverage, we also remove a foreground-contaminated \enquote{wedge} \citep{Morales2012, Vedantham2012, Trott2012, Parsons2014, Pober2014, Liu2014a, Liu2014b, Murray2018, Liu2020}.  Conservatively, we remove (zero) all modes below the horizon limit, which can be expressed as a slope in the line-of-sight ($k_{\|}$) vs. sky-plane ($k_\perp$) Fourier modes:
\begin{equation}
    k_{\|} \le k_\perp \frac{E(z)}{1+z} \int_0^z \frac{dz'}{E(z')} \, ,
    \label{eq:horizon_cut_condition}
\end{equation}
where $E(z)=\sqrt{\Omega_m (1+z)^3 + \Omega_\Lambda}$. For more details on the telescope noise and foreground avoidance implemented in this work, we refer reader to \cite{prelogovic_22}. The only difference with respect to the original method is that we do not apply a \enquote{rolling} of the wedge filter, more relevant for 21-cm images. As the 1D PS used here is computed in (binned) Fourier space, it is sufficient to apply the wedge filter once per the lightcone chunk on which the PS is calculated.

After the telescope effects are included, we cut the resulting 21-cm brightness temperature lightcone, $\delta {T}_b(\bm{x}, z)$, along redshift axis into chunks of $300 \, \mathrm{Mpc}$ and compute the 1D PS on each section as:
\begin{equation}
    \delta\bar{T}^2_b \Delta_{21}^2(k, z) \equiv \frac{k^3}{2 \pi^2 V} \left\langle\left|\delta T_b(\bm{k}, z) - \delta\bar{T}_b(z)\right|^2\right\rangle_k \, ,
    \label{eq:1DPS}
\end{equation}
where $z$ is the central redshift of each chunk.  This 1D PS serves as our summary statistic throughout this work.

\subsection{Model parameters} \label{subs:model_parameters}

We characterize the unknown UV and X-ray properties of high-$z$ galaxies using the model from \citet{Park_2019}.  The simple scaling relations from that work are both easy to interpret (thus allowing us to set well motivated priors), and sufficient to recover currently-available galaxy and EoR observations. Specifically, we use a subset of five parameters from \citet{Park_2019} that drive most of the variation in the signal:
\begin{itemize}
    \item $\bm{f_{*, 10}}$ - fraction of the galactic gas in stars, normalized at a halo mass of $10^{10} M_\odot$. The stellar mass of a galaxy $M_*$ is assumed to be proportional to the halo mass $M_h$ as $$M_* = f_* \left(\frac{\Omega_b}{\Omega_m}\right) M_h \, ,$$ where $f_*$ is a power-law $$f_*(M_h) = f_{*, 10} \left(\frac{M_h}{10^{10} M_\odot}\right)^{\alpha_*} \, .$$
    Here, we fix the power index to $\alpha_* = 0.5$, which fits the observed, faint-end\footnote{The general stellar to halo mass relation is better fit with a double power law, with a flatter index for halos above $\sim10^{12} M_\odot$ (e.g. \citealt{Mirocha_2019, Behroozi_2019}).  However, the radiation fields that drive the 21-cm signal are dominated by the far more abundant faint galaxies (e.g. \citealt{Bouwens_2015}, Nikolić et al. in prep); thus we do not bother to characterize the rare massive galaxies (or AGN) that have a negligible contribution to cosmic radiation fields.}
    UV luminosity functions very well (see \citealt{Park_2019} and references therein).
    \item $\bm{f_{\text{esc}, 10}}$ - the ionizing UV escape fraction, normalized at a halo mass of $10^{10} M_\odot$. The ionizing escape fraction is taken to scale with the halo mass as a power law $$f_{\text{esc}}(M_h) = f_{\text{esc}, 10} \left(\frac{M_h}{10^{10} M_\odot}\right)^{\alpha_{\text{esc}}} \, .$$ Here again we fix the power law index to $\alpha_{\text{esc}} = -0.5$ (c.f. \citealt{Park_2019}).
    \item $\bm{M_{\text{turn}}}$ - a charactisitic halo mass below which star formation becomes inefficient due to slow gas accretion, photoheating and/or supernovae feedback.  We assume only a fraction of halos given by $$f_{\text{duty}}(M_h) = \exp\left(-\frac{M_{\text{turn}}}{M_h}\right) \, .$$ can host star forming galaxies (c.f. \citealt{OShea_2015, Xu_2016, Sobacchi2013, Sobacchi2013b}).
    \item $\bm{E_0}$ - the energy threshold below which X-ray photons are absorbed by the ISM of the host galaxy. This will depend on the ISM density, metallicity, as well as the local environment of X-ray sources (like high mass X-ray binary stars; e.g. \citealt{Das_2017}).
    \item $\bm{L_{X < 2\mathrm{keV}} / \mathrm{SFR}}$ - the soft-band (with energies between $E_0$ -- 2 keV) X-ray luminosity per unit star formation escaping the galaxy, in units of $\mathrm{erg \, s^{-1} \, keV^{-1} \, M_{\odot}^{-1} \, yr}$.
    Photons with energies greater than $2\mathrm{keV}$ have mean free paths larger than the Hubble length during the CD (e.g. \citealt{McQuinn2012}), while we assume photons with energies below $E_0$ are absorbed in the host galaxy. We assume a SFR of: $\dot{M}_* = M_*/(t_* H^{-1})$, where $t_*$ controls the characteristic timescale as a fraction of Hubble time, $H(z)$. Here, we fix it to $t_* = 0.5$ (c.f. \citealt{Park_2019}).
\end{itemize}

Our databases are created by sampling physically-motivated priors for the above five galaxy parameters.  We take flat priors for X-ray parameters $E_0 / [\mathrm{keV}] \in [0.2, 1.5]$ and $\log_{10} \frac{L_{X<2\mathrm{keV}} / \text{SFR}} {[\mathrm{erg \, s^{-1} \, keV^{-1} \, M_{\odot}^{-1} \, yr}]} \in [38, 42]$, as they are only weakly and indirectly constrained for the high-mass X-ray binaries that likely dominate X-ray production in the first galaxies (e.g. \citealt{Mineo2012, Fragos2013, Lehmer_2016, Das_2017}). Our priors for star formation and UV emission are taken from the posterior of \cite{Park_2019}, constructed from existing measurements of (i) UV luminosity functions \citep{Bouwens2015b, Bouwens2017, Oesch2018}, (ii) $\tau_e$ \citep{Planck_2016} and (iii) the dark fraction from the Lyman forest \citep{McGreer2015}. Specifically, We fit the 6D posterior distribution from \citet{Park_2019} with a kernel density estimator, and evaluate it at the fiducial values of ($\alpha_*=0.5$, $t_*=0.5$ and $\alpha_{\text{esc}}=-0.5$) to obtain our prior over $f_{*, 10}$, $f_{\text{esc}, 10}$ and $M_{\text{turn}}$.  For this we use the \texttt{conditional\_kde}\footnote{\url{https://github.com/dprelogo/conditional_kde}} code.  We stress that the posteriors that we obtain from the 21-cm PS are much narrower than the priors, which means that the inference is likelihood-dominated and not sensitive to our specific choice of priors.

When required below, we assume a fiducial parameter set, $\bm{\theta}_{\rm fid}$ consistent with \cite{Park_2019}: $\log_{10} f_{*, 10} = -1.3$, $\log_{10} f_{\text{esc}, 10} = -1.0$, $\log_{10} M_{\text{turn}} / M_{\odot} = 8.7$, $E_0 = 0.5 \mathrm{keV}$, $\log_{10} \frac{L_{X<2\mathrm{keV}} / \text{SFR}} {[\mathrm{erg \, s^{-1} \, keV^{-1} \, M_{\odot}^{-1} \, yr}]} = 40$. For ease of notation, in the remainder of the paper we drop the $\log_{10}$, units and limits, denoting the parameters as: $f_{*, 10}$, $f_{\text{esc}, 10}$, $E_0$, $M_{\text{turn}}$, $L_X/ \text{SFR}$.

\section{Choosing a likelihood function} \label{ch:choosing_a_likelihood}

In this work our summary statistic (i.e. our \enquote{data space}) is the 1D power spectrum from Equation \ref{eq:1DPS}. Here we compare different choices for the likelihood (both explicit and implicit).

\subsection{Classical (explicit) Gaussian likelihoods}
By far the most common choice for the 21-cm PS likelihood is a Gaussian with a fixed variance (i.e. a diagonal covariance matrix in $(k, z)$ space) computed at some fiducial value $\bm{\theta}_{\rm fid}$, as in Equation~\ref{eq:gaussian_likelihood}.  When doing a 21-cm forecast, this $\bm{\theta}_{\rm fid}$ is chosen a priori to make a mock observation.  More generally, one can use a maximum likelihood estimate or a maximum a posteriori (MAP) from inference using other data sets to choose $\bm{\theta}_{\rm fid}$.

Here we discuss all of the ingredients for computing a Gaussian likelihood. For every point in the parameter space $\bm{\theta}$, we label the Gaussian mean, variance, or full covariance as $\bm{\mu}(\bm{\theta})$, $\bm{\sigma}^2(\bm{\theta})$, or $\Sigma(\bm{\theta})$, respectively. In the case of the fiducial parameters $\bm{\theta}_{\text{fid}}$, we calculate them in the following way:
\begin{align}
    \bm{\mu}_{k_i, z_j}(\bm{\theta}_{\text{fid}}) &= \frac{1}{N} \sum_{n = 1}^N  \Delta^2_{21}(\bm{\theta}_{\text{fid}}, k_i, z_j)_n \, , \label{eq:gaussian_mean}\\
    \Sigma_{k_i, z_j, k_l, z_m}(\bm{\theta}_{\text{fid}}) &= \frac{1}{N - 1} \sum_{n=1}^N (\Delta^2_{21}(\bm{\theta}_{\text{fid}}, k_i, z_j)_n - \bm{\mu}_{k_i, z_j}) \times \nonumber \\
    &\times (\Delta^2_{21}(\bm{\theta}_{\text{fid}}, k_l, z_m)_n - \bm{\mu}_{k_l, z_m}) \, , \label{eq:gaussian_cov}\\
    \bm{\sigma}_{k_i, z_j}^2(\bm{\theta}_{\text{fid}}) &= \Sigma_{k_i, z_j, k_i, z_j}(\bm{\theta}_{\text{fid}}) \, , \label{eq:gaussian_var}
\end{align}
where $n$ represents a different realization of the cosmic random seed (i.e. different initial conditions) and $k_i, z_j$ are Fourier wavemode and redshift bins, respectively. We will see later how to estimate these quantities using machine learning. Note that there are two main sources of stochasticity in our pipeline: the cosmological signal and the telescope noise. The telescope noise does not depend on $\bm{\theta}$ and can be averaged over several realizations and pre-computed. Therefore, the total mean is estimated by: $\bm{\mu}(\bm{\theta}) = \bm{\mu}_{21\mathrm{cm}}(\bm{\theta}) + \bm{\mu}_{\text{noise}}$. The contributions are separable even with the wedge excision.

Finally, we can write the Gaussian likelihood for an observed power spectrum, $\Delta^2_{21 \, \text{obs}}$, in standard notation as:
\begin{equation}
    \mathcal{L}(\Delta^2_{21 \, \text{obs}} | \bm{\theta}) = \mathcal{N}(\Delta^2_{21 \, \text{obs}}| \bm{\mu}(\bm{\theta}), \Sigma(\bm{\theta})) \, . 
    \label{eq:classic_likelihood_cov}
\end{equation}    
Here, $\Sigma(\bm{\theta})$ is the covariance 
and $\bm{\mu}(\bm{\theta})$ the PS mean, estimated at each $\bm{\theta}$.
The vast majority of 21-cm literature also ignores correlations between different redshifts and $k$-modes:
\begin{equation}
    \mathcal{L}(\Delta^2_{21 \, \text{obs}} | \bm{\theta}) = \mathcal{N}(\Delta^2_{21 \, \text{obs}}| \bm{\mu}(\bm{\theta}), \bm{\sigma}^2(\bm{\theta})) \, ,
    \label{eq:classic_likelihood_var}
\end{equation} 
where the variance $\bm{\sigma}^2(\bm{\theta})$ is simply the diagonal of the covariance (Eq. \ref{eq:gaussian_var}).

To correctly estimate the mean and (co)variance of the forward-modeled power spectra, one would need to run several simulations changing the cosmic seed (c.f. Eq. \ref{eq:gaussian_mean}) for each $\bm{\theta}$ sample. However for computational reasons, the cosmic seed is usually not varied when performing inference and thus {\it the mean is estimated from only one, fixed realization} (though techniques such as "pairing and fixing" could reduce the computation time; e.g. \citealt{Giri2023}).  Moreover, the cosmic (co)variance is generally only computed at the fiducial parameter values (used to make the mock observation), i.e. constant $\Sigma(\bm{\theta}_{\rm fid})$ and $\bm{\sigma}^2(\bm{\theta}_{\rm fid})$ are used in place of the $\bm{\theta}$-dependent (co)variances in equations (\ref{eq:classic_likelihood_cov}) -- (\ref{eq:classic_likelihood_var}).   Both of these assumptions can bias inference \citep{Mondal2017, Shaw2019, Watkinson_2022, Greig_2023}.

Here when performing \enquote{classical} (non-amortized) inference with the explicit likelihoods given by equations (\ref{eq:classic_likelihood_cov}) and (\ref{eq:classic_likelihood_var}), we pick a \textit{model seed} which gives a realization of the 21-cm PS that is {\it closest} (among 1000 realizations) to the mean at $\bm{\theta}_{\rm fid}$, and then use that same cosmic seed when sampling all of the parameter space $\bm{\theta}$.  Note that this is an improvement over the most common approach of {\it randomly} choosing the model seed and assuming it is a proxy for the mean.
Furthermore for $\Delta^2_{21 \, \text{obs}}$, we create a mock observation at $\bm{\theta}_{\text{fid}}$ by picking a \enquote{reasonable} \textit{data seed} that is within $1 \, \sigma$ of the mean.  Finally, for computational reasons, we compute the variance and covariance from 1000 realizations only at the fiducial values, $\bm{\theta}_{\rm fid}$, keeping these uncertainties constant throughout parameter space.

\begin{figure*}
    \centering
    \subfigure{\includegraphics[width=.4\linewidth]{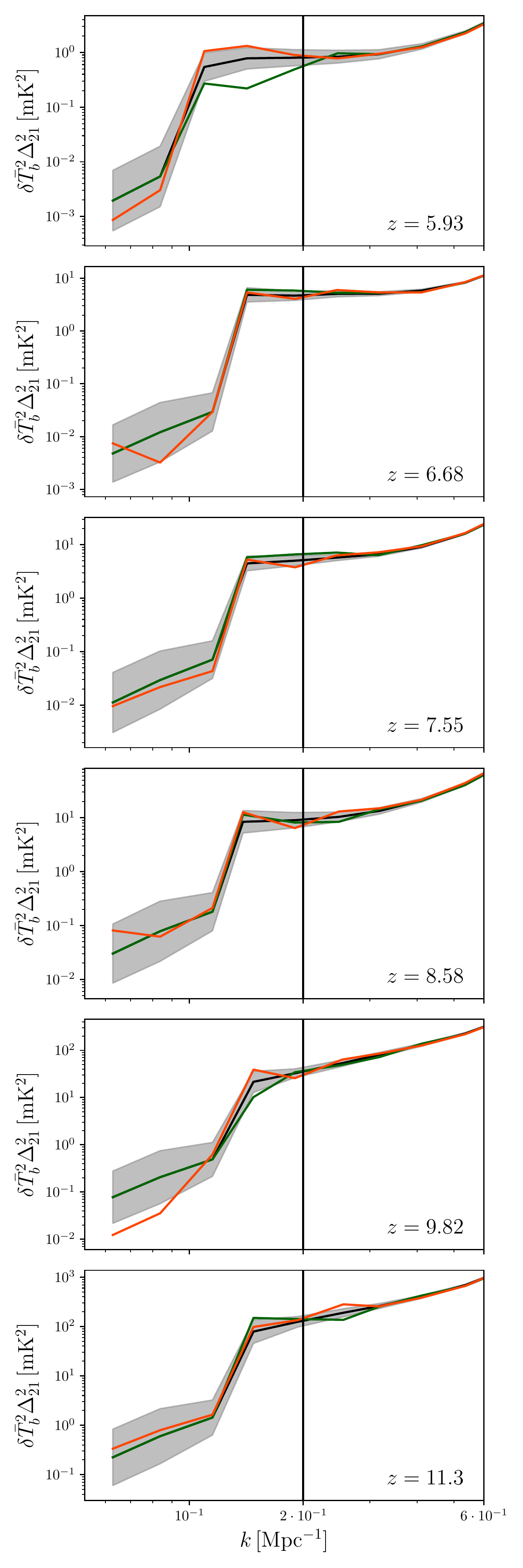}}
    \subfigure{\includegraphics[width=.4\linewidth]{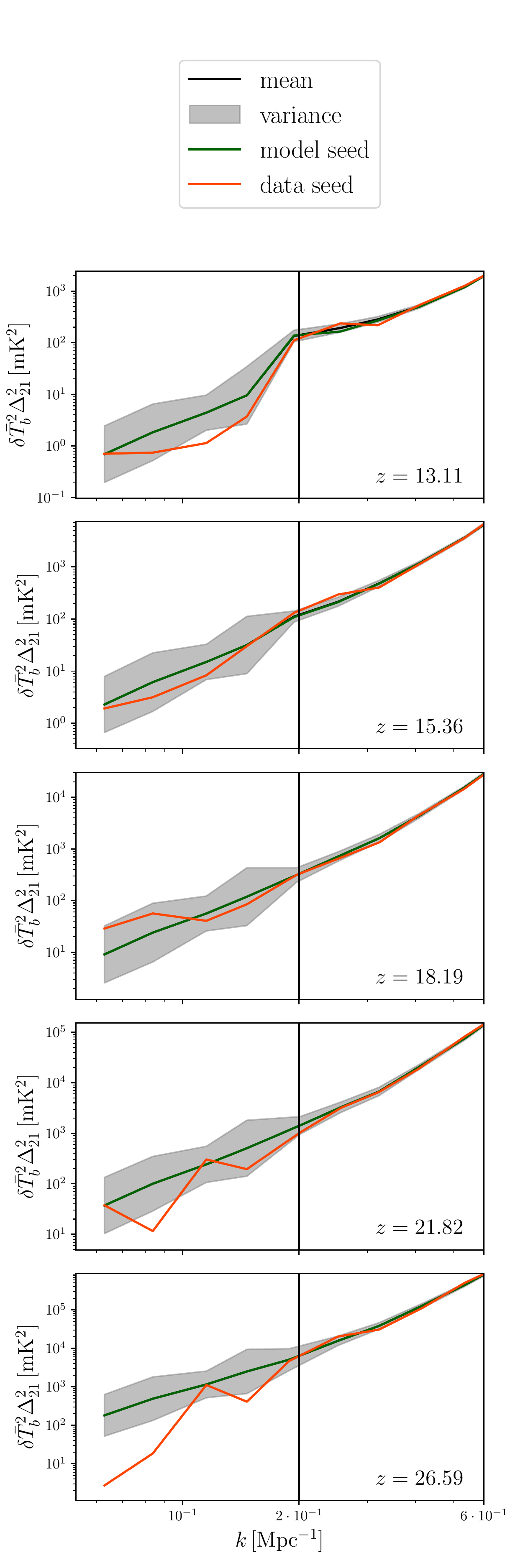}}
    \caption{The fiducial power spectrum mean ({\it black curves}) and standard deviation ({\it grey region}), computed from $1000$ realizations varying initial conditions and noise seeds at $\bm{\theta}_{\rm fid}$.  We note that the power spectrum includes contributions from the cosmic signal, telescope noise and wedge excision; c.f. Fig. \ref{fig:pipeline}.  The \enquote{data seed} used as our mock observation is shown in red and is chosen to be a statistically \enquote{reasonable} realization. The \enquote{model seed} is shown in green and is chosen to be the realization closest to the mean, so that we can avoid varying the initial conditions when performing classical inference.  The sharp drop seen at low $k$ results from the wedge excision. All $k$-modes to the right of the vertical line $k \ge 0.2 \mathrm{Mpc}^{-1}$ are taken into account when computing the likelihood.}
    \label{fig:variance_likelihood}
\end{figure*}

The resulting mean PS, its variance, as well as the realizations corresponding to the model and data seeds are shown in Figure \ref{fig:variance_likelihood}, for the fiducial parameter set $\bm{\theta}_{\rm fid}$. Due to the wedge excision, we consider only $k \ge 0.2 \mathrm{Mpc}^{-1}$ in our analysis, marked with vertical line. The upper limit $k \le 0.6 \mathrm{Mpc}^{-1}$ is motivated by the drop in $uv$ coverage as one approaches the maximum core baseline (see \citealt{prelogovic_22} for more details). We see from the figure that our choice of model seed matches the mean almost perfectly, at least for this fiducial parameter value, $\bm{\theta}_{\rm fid}$.  This is a computationally inexpensive way of approximating the actual mean that should be computed on-the-fly when using a Gaussian likelihood for classical inference (c.f. Equation \ref{eq:gaussian_likelihood}).  Furthermore, we verify that our data seed used for the mock observation is \enquote{reasonable}, and does not correspond to a rare outlier.

\begin{figure*}
    \centering
    \includegraphics[width=.99\linewidth]{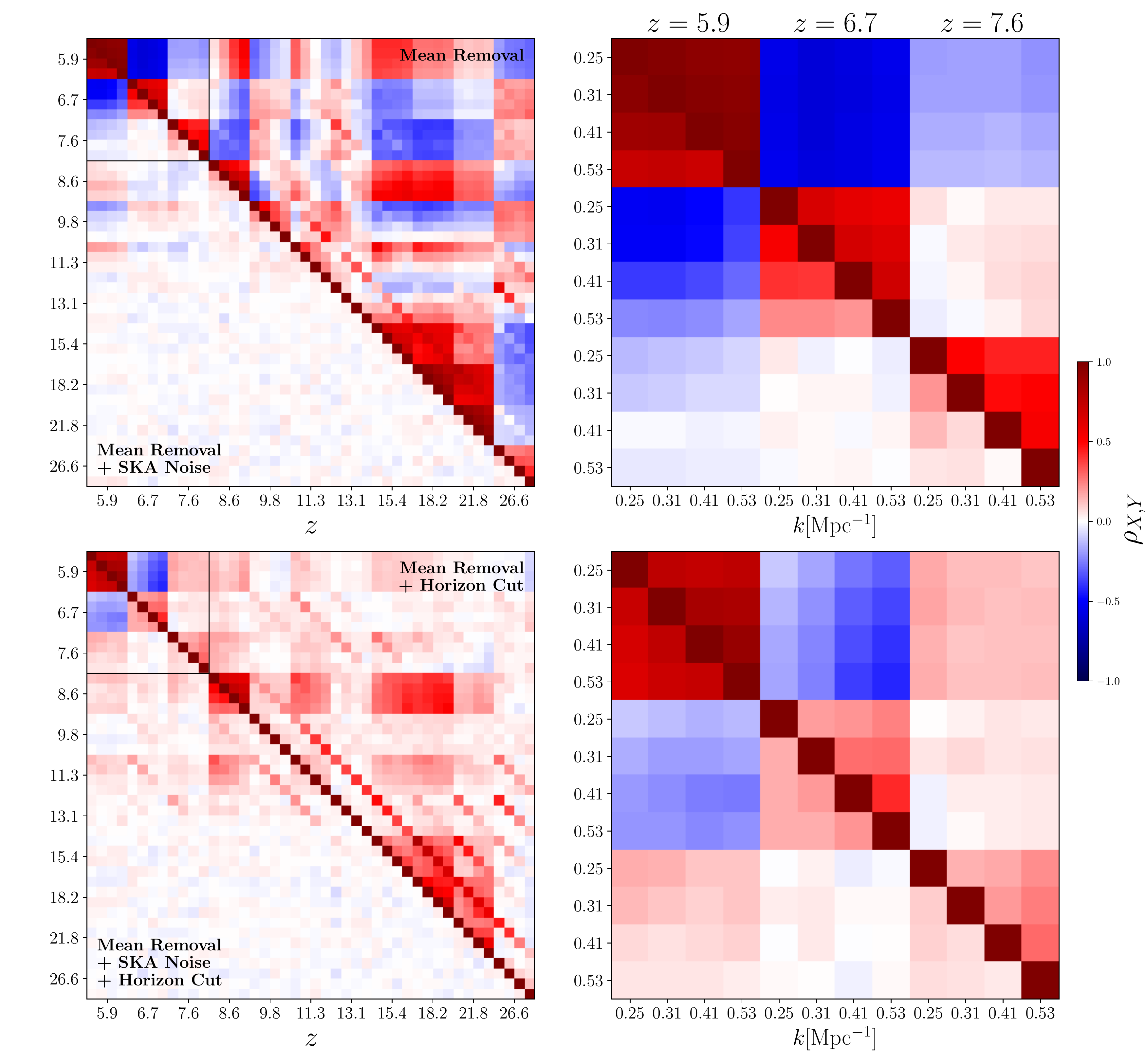}
    \caption{Correlation matrices of the 21-cm PS at the fiducial parameter choice, $\bm{\theta}_{\rm fid}$ (c.f. Eq. \ref{eq:gaussian_cov}). \textbf{Upper-left}: flattened 4D correlation matrix for the \textit{Mean Removal} case with and without SKA noise ({\it lower left} and {\it upper right} triangles, respectively). \textbf{Upper-right}: zoom-in of the upper-left plot for the first three redshift bins. \textbf{Bottom row}: the same as the upper row, with addition of  \textit{+ Horizon Cut}.}
    \label{fig:corr_matrices}
\end{figure*}

 In Figure \ref{fig:corr_matrices} we show the correlation matrix of the 21-cm PS computed at the fiducial parameters, $\Sigma_{k_i, z_j, j_l, z_m}(\bm{\theta}_{\rm fid})$ (c.f. Equation \ref{eq:gaussian_cov}). We calculate it on $11$ distinct lightcone chunks and $4$ logarithmically-spaced k-bins, in the range $0.2 - 0.6 \, \mathrm{Mpc^{-1}}$. To visualize such a 4D correlation matrix $(k_i, z_j, k_l, z_m)$, one has to flatten it to 2D, which is shown in the top-left. There are $11$ distinct redshift blocks, with $4\times4$ k-bins. The image is further split into two triangles. The upper right half shows correlations after \textit{Mean Removal}, and the lower left half shows correlations after \textit{Mean Removal + SKA Noise} (c.f. Figure \ref{fig:pipeline}). The upper right panel shows a zoom-in on the first three redshift chunks of the upper left panel. The bottom row shows the same quantities but including also \enquote{+ Horizon Cut}.

As expected, the diagonal terms (the variance) have the strongest correlation, and the telescope noise strongly suppresses off-diagonal correlations above $z \gtrsim 10$.  However, the off-diagonal terms can be significant at $z \lesssim 10$, suggesting that it is important to include the full covariance matrix in the Gaussian likelihood for inference during the epoch of reionization (EoR).
 Similar conclusions have also been reached by \citet{Mondal2017, Shaw2019} (see their Fig. 6 and Fig. 7, respectively), where the authors analyzed correlations between $k$-bins at fixed redshifts. 
 
\begin{figure}
    \centering
    \includegraphics[width=.99\linewidth]{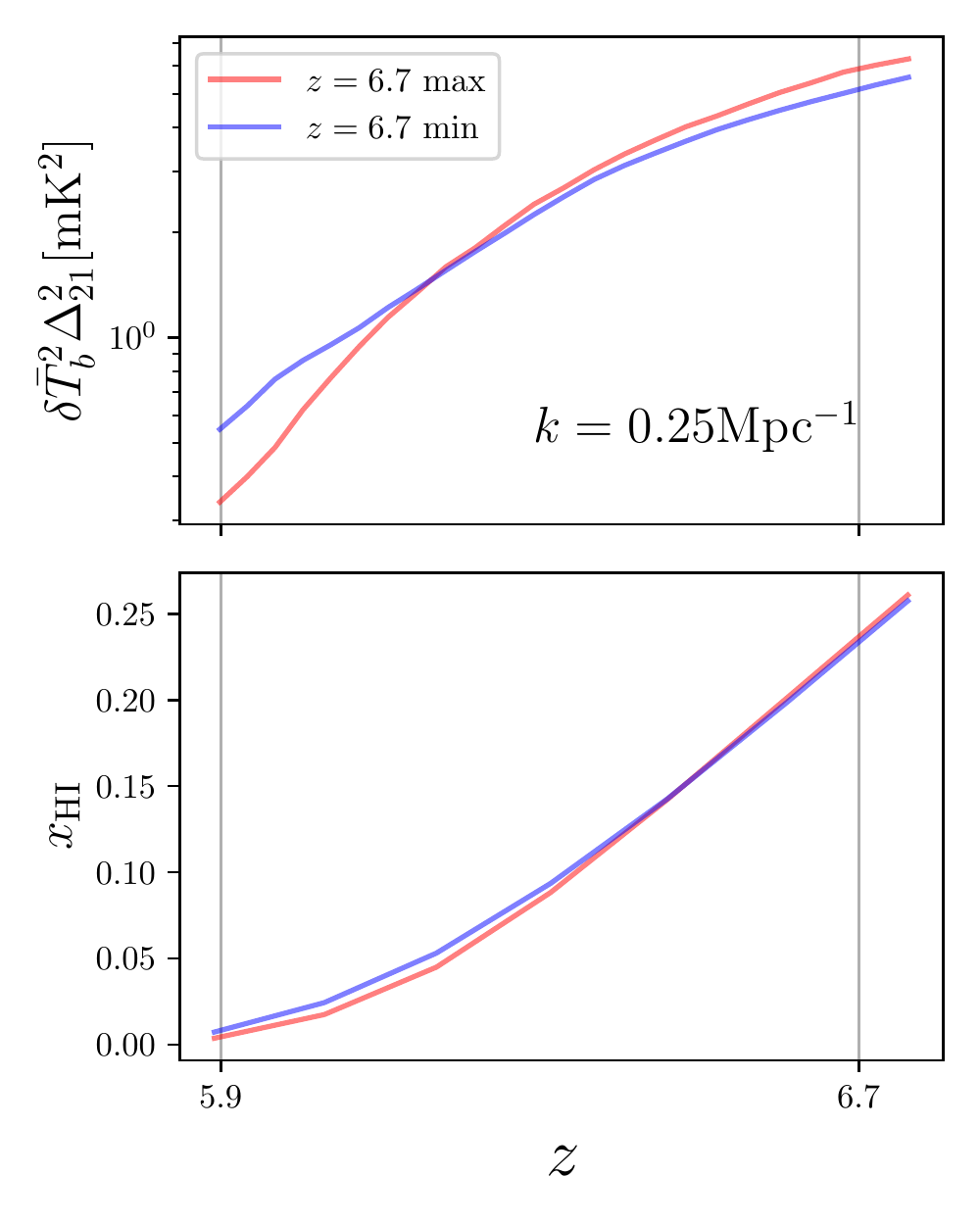}
    \caption{Redshift evolution of the 1D PS at $k =  0.25 \mathrm{Mpc^{-1}}$ ({\it upper panel}) and IGM mean neutral fraction ({\it lower panel}), covering the first two redshift bins. The red (blue) curves illustrate the mean computed over 5 realizations with large (small) power at $z = 6.7$.  Consistent with Figure \ref{fig:corr_matrices}, the panels demonstrate strong anti-correlation between $z = 5.9$ and $z = 6.7$, resulting from the relative differences in the steepness in the halo mass functions of these realizations, which impact the timing of the EoR (see text for details). }
    \label{fig:explaining_anticorrelation}
\end{figure}

In order to gain physical intuition about the origin of the strongest (anti-)correlations, in Figure \ref{fig:explaining_anticorrelation} we show the redshift evolution of the PS amplitude at $k = 0.25 \mathrm{Mpc^{-1}}$ during the EoR (upper panel), together with the neutral fraction (lower panel).  The blue (red) line shows the mean computed over several realizations that have high (low) power at $z = 6.7$.  One can clearly see that those cosmic seeds that result in high 21-cm power at $z = 6.7$ subsequently have low power at $z = 5.9$, consistent with the (anti-)correlation redshift lengths seen in Figure \ref{fig:corr_matrices}.  In other words, the realizations in blue have a delayed, but subsequently more rapid EoR evolution, seen in the lower panel.  This is something that is expected if the EoR had a larger contribution from rare, massive halos that are on the exponential tail of the halo mass function.  These massive halos appear later than their lighter counterparts, but their fractional abundance subsequently increases more rapidly (e.g. \citealt{barkana2004}).
We confirm this explanation by looking at the PDFs of the Lagrangian densities of the realizations in red vs. those in blue.  Indeed, the realizations shown in blue have a larger, high-sigma tail in the density PDF: these rare, dense cells increase the relative abundance of the most massive halos (flattening the halo mass function), resulting in a delayed and more rapid EoR evolution.


In what follows, we will be fitting the likelihood with different density estimators. For this, it helps to consider the likelihood as a function in the parameter space $\bm{\theta}$ and a {\it probability density} in the power spectrum {\it data space}, $\bm{d}_{PS}$. Using this notation, a Gaussian likelihood can be written as:
\begin{equation}
    \mathcal{L}(\bm{d}_{PS} | \bm{\theta}) = \mathcal{N}(\bm{d}_{PS}| \bm{\mu}(\bm{\theta}), \Sigma(\bm{\theta})) \, . 
\end{equation}
Therefore, $\mathcal{L}(\Delta^2_{21 \,  \text{obs}}| \bm{\theta})$  from Eqs. \ref{eq:classic_likelihood_cov}, \ref{eq:classic_likelihood_var} is a function of $\bm{\theta}$ for a {\it given} observation $\Delta^2_{21 \,  \text{obs}} \in \bm{d}_{PS}$,  while $\mathcal{L}(\bm{d}_{PS} | \bm{\theta})$ is a probability density over all possible observations $\bm{d}_{PS}$.

\subsection{Simulation-based (\enquote{likelihood free}) inference with neural density estimators}
\label{subs:SBI_with_NDE}
In the previous section we introduced two classic likelihood choices used in the vast majority of 21-cm inference.  These explicit, Gaussian likelihoods are only approximations to the true likelihood for several reasons:
\begin{enumerate}
    \item the mean of the forward-modeled power spectra is estimated from one simulation, instead of co-varying the initial condition (IC) random seeds to compute $\bm{\mu}(\bm{\theta}) = \langle \Delta_{21}^2(\bm{\theta}) \rangle_{\rm ICs}$ at each parameter sample $\bm{\theta}$ (see Eq. \ref{eq:gaussian_mean}),
    \item the (co)variance is generally computed only at $\bm{\theta}_\text{fid}$, instead of computing it at each parameter sample $\bm{\theta}$,
    \item only 1-pt and 2-pt correlations between wavemodes and redshifts are included;  higher order terms are ignored.
\end{enumerate}
In the last point, we refer to higher order moments between wavemodes and redshifts used in the PS likelihood, {\it not} the fact that the PS itself is a two point statistic and so is a sub-optimal summary statistic for non-Gaussian fields.  We will quantify the later point in future work (Prelogović et al. in prep).
With the classical methods above, relaxing any of these simplifications seriously increases the computational costs, often making non-amortized inference impractically expensive. 

Simulation-based inference offers an efficient way to address all of the above limitations, by using a simulator to draw samples from the \enquote{real} likelihood. As discussed in the introduction, this involves sampling all of the relevant sources of stochasticity in the simulator (in our case the initial conditions and thermal noise), together with the astrophysical parameters, so as to create a training set of mock observations (i.e. a joint distribution of $P(\bm{d}_{PS}, \bm{\theta})$).

There are many different flavors of SBI, depending on the usage-case. Here we use density estimation likelihood-free inference (DELFI, \citealt{Papamakarios_2018, Alsing_2018, Alsing_2019}), which uses a density estimator to fit either the likelihood $\mathcal{L}(\bm{d}_{PS} | \bm{\theta})$ or the posterior $P(\bm{\theta} | \bm{d}_{PS})$. We choose the former, which allows us to parametrize the form of the likelihood in some of our investigations, directly testing the impact of the previously mentioned approximations.  Moreover, fitting the likelihood is more flexible as the priors can be changed afterwards without having to retrain. We can then perform inference using an explicit or non-explicit  likelihood fitted with density estimation.  This inference is then vastly more computationally efficient as the simulator (in our case \cmfast) is not called at each likelihood evaluation.  This is why SBI is a form of so-called {\it amortized} inference: the computational cost is upfront in generating a training set for the density estimation.  Once the likelihood is learned via density estimation, subsequent inference is done very rapidly, with a single likelihood evaluation taking of order miliseconds. 

Another and more popular form of amortized inference in 21-cm literature is emulation, where a deep neural network is trained to directly replace the simulator for pre-chosen summary statistics like the PS  (e.g.  \citealt{emuPSKern17, emuPSSP17, emuPSinvS17, emuPSJ19, emuPSG20, emuPSM21}, Breitman et al. in prep.).  However, emulators themselves have errors which are difficult to include properly in the likelihood.  With SBI, our only source of error (assuming the simulator is correct) is the density estimation, which can always be reduced by increasing the training set or using iterative re-sampling.

\subsubsection{NDE loss function}
Given a NN parametrization of a likelihood $\mathcal{L}_\mathrm{NN}(\bm{d}_{PS} | \bm{\theta})$, one can train its parameters to approximate the real likelihood $\mathcal{L}(\bm{d}_{PS} | \bm{\theta})$. A natural distance between these distributions is the Kullback-Liebler (KL) divergence. For a continuous 1D random variable, with $p(x)$ as the true (target) distribution and $q(x)$ as the approximate distribution, the integral
\begin{equation}
    D_{\mathrm{KL}}(P \| Q) = \int_{-\infty}^{\infty} p(x) \log\frac{p(x)}{q(x)} \, \mathrm{d}x
\end{equation}
represents the information gain if the target distribution $p(x)$ were used instead of $q(x)$. Therefore, the main goal during training is to minimize this difference. The integral can be further written as:
\begin{align}
    D_{\mathrm{KL}}(P \| Q) &=  \int_{-\infty}^{\infty} p(x) \log p(x) \, \mathrm{d}x - \int_{-\infty}^{\infty} p(x) \log q(x) \, \mathrm{d}x \nonumber\\
    &\approx -H_P - \frac{1}{N}\sum_{i=1}^{N} \log q(x_i) \, ,
    \label{eq:KL_divergence_explicit}
\end{align}
with samples $x_i$ being drawn from $p(x)$. In the last expression, $H_P$ is a constant denoting the Shannon entropy of the distribution $p(x)$ and the second term is the Monte-Carlo estimation of the integral $\int_{-\infty}^{\infty} p(x) \log q(x) \, \mathrm{d}x$.

We use Eq. (\ref{eq:KL_divergence_explicit}) to define the loss between the NN likelihood estimate, $\mathcal{L}_\mathrm{NN}(\bm{d}_{PS} | \bm{\theta}) \equiv q(x)$ and the true likelihood $\mathcal{L}(\bm{d}_{PS} | \bm{\theta}) \equiv p(x)$:
\begin{equation}
    L =  - \frac{1}{N}\sum_{i=1}^{N} \log \mathcal{L}_{\mathrm{NN}} (\bm{d}_{PS, i} | \bm{\theta}_i) \, ,
    \label{eq:loss_function}
\end{equation}
where $\bm{\theta}_i, \bm{d}_{PS, i}$ are samples from the training set.
Once the parametric shape of $\mathcal{L}_{\mathrm{NN}}$ is defined (see the following section), one can minimize the loss via stochastic gradient descent (SGD) and back propagation.\footnote{In this work, we sample parameters from the prior $\pi(\bm{\theta})$ and then draw from the likelihood $\mathcal{L}(\bm{d}_{PS} | \bm{\theta})$ by simulating the PS. In this way, $\bm{\theta}_i, \bm{d}_{PS, i}$ are sampled from the joint distribution $P(\bm{\theta}, \bm{d}_{PS})$. One could therefore think of defining the loss function as  $$L =  - \frac{1}{N}\sum_{i=1}^{N} \pi(\bm{\theta}_i)^{-1} \log \mathcal{L}_{\mathrm{NN}} (\bm{d}_{PS, i} | \bm{\theta}_i) \, .$$ As we are interested in the distribution conditional on the parameter space (i.e. likelihood), both methods are correct. Introducing the factor $\pi(\bm{\theta}_i)^{-1}$ in the loss gives more weight to the samples from the prior edges, counter-balancing the fact that they are rare. We prefer to focus on the main prior volume, so we keep the loss as defined in the main text.}

In this work, we pre-calculate a database of $\approx 78\,000$ lightcones, pass them through the observational pipeline and compress to the power spectra (cf. Figure \ref{fig:pipeline}). Realizations correspond to astrophysical parameters drawn from the prior distribution described in \S\ref{subs:model_parameters}. Importantly, every realization has different cosmic and noise seeds. Around $ 57\,000$ are used for the training, $14\,000$ for validation and $7\,000$ for the final testing.  We note that a database of this size corresponds roughly to the number of samples (likelihood calls) needed for MCMC or nested inference to converge.  We return to this point below, stressing that SBI is not only more accurate but also more computationally efficient compared to classical inference when the simulator is called on-the-fly (non-amortized).

\subsubsection{NDE likelihood parametrizations}
We use SBI to test the  validity of the common approximations made when using a Gaussian likelihood. As discussed at the beginning of this section (\S\ref{subs:SBI_with_NDE}), these include: 
\begin{enumerate}
    \item estimating the mean from a single realization,
    \item estimating the (co)variance only at the fiducial parameters,
    \item assuming a Gaussian form to begin with, thus ignoring higher order correlations.
\end{enumerate}
Our aim is to investigate the consequences of each assumption, quantifying their impact on recovered posteriors. To do that, we relax constraints one by one, gradually enlarging the likelihood complexity and possibly improving final performance. For this, we use neural density estimators parameterized with simple fully-connected layers. 

In order to estimate the mean of the PS better, we use a feed-forward NN which takes parameters $\bm{\theta}$ and outputs the PS mean. In other words,
\begin{equation}
    \bm{\mu}_{\text{NN}}(\bm{\theta}) = \text{NN}(\bm{\theta}) \, .
\end{equation}
The possible Gaussian likelihoods are then:
\begin{align}
    \mathcal{L}_{\text{NN}}(\bm{d}_{PS} | \bm{\theta}) &= \mathcal{N}(\bm{d}_{PS}| \bm{\mu}_{\text{NN}}(\bm{\theta}), \bm{\sigma}^2(\bm{\theta}_{\text{fid}})) \, , \\
    \mathcal{L}_{\text{NN}}(\bm{d}_{PS} | \bm{\theta}) &= \mathcal{N}(\bm{d}_{PS}| \bm{\mu}_{\text{NN}}(\bm{\theta}), \Sigma(\bm{\theta}_{\text{fid}})) \, .
\end{align}
Here $\bm{\sigma}^2(\bm{\theta}_{\text{fid}})$ and $\Sigma(\bm{\theta}_{\text{fid}})$ represent the variance and covariance estimated at the fiducial parameter values, respectively. After training such likelihoods, the network will learn to interpolate PS mean values for the whole parameter space $\bm{\theta}$.

Likewise, we can also estimate the (co)variance matrix with a NN. In this scenario, the network can output one of the following:
\begin{align}
    \bm{\mu}_{\text{NN}}(\bm{\theta}), \bm{\sigma}^2_{\text{NN}}(\bm{\theta}) = \text{NN}(\bm{\theta}) \, , \\
    \bm{\mu}_{\text{NN}}(\bm{\theta}), \Sigma_{\text{NN}}(\bm{\theta}) = \text{NN}(\bm{\theta}) \, ,
\end{align}
with their respective likelihoods:
\begin{align}
    \mathcal{L}_{\text{NN}}(\bm{d}_{PS} | \bm{\theta}) &= \mathcal{N}(\bm{d}_{PS}| \bm{\mu}_{\text{NN}}(\bm{\theta}), \bm{\sigma}^2_{\text{NN}}(\bm{\theta})) \, , \\
    \mathcal{L}_{\text{NN}}(\bm{d}_{PS} | \bm{\theta}) &= \mathcal{N}(\bm{d}_{PS}| \bm{\mu}_{\text{NN}}(\bm{\theta}), \Sigma_{\text{NN}}(\bm{\theta})) \, .
\end{align}
The last equation represents a likelihood complexity currently intractable by classical methods, as calculating a full covariance matrix at every point of the parameter space is too expensive. 

Finally, we also adopt a non-explicit likelihood, allowing us to test the importance of the higher-order terms that are ignored when assuming a Gaussian form.  We try two non-explicit likelihood density estimators implemented in DELFI: (i) a  Gaussian mixture network; and (ii) conditional masked autoregressive flows (CMAF).  A Gaussian mixture density network is a simple density estimator that is very stable to train.\footnote{A Gaussian mixture density network should not be confused with a Gaussian mixture model. The latter describes a distribution where each Gaussian represents samples of a particular class. This is not the case here - the mixture density network is used purely for better parametrization of the likelihood.} For a mixture of $K$ Gaussians, a feed-forward NN takes parameters $\bm{\theta}$ and outputs
\begin{equation}
    \bm{\mu}_{\text{NN}, 1}(\bm{\theta}), \Sigma_{\text{NN}, 1}(\bm{\theta}), \ldots, \bm{\mu}_{\text{NN}, K}(\bm{\theta}), \Sigma_{\text{NN}, K}(\bm{\theta}), \Phi(\bm{\theta}) = \text{NN}(\bm{\theta}) \, ,
\end{equation}
where $\bm{\mu}_{\text{NN}, i}(\bm{\theta}), \Sigma_{\text{NN}, i}(\bm{\theta})$ describe mean and covariance of the $i-\text{th}$ Gaussian and $\phi_i(\bm{\theta}) \in {\Phi(\bm{\theta})}$ its relative weight, where $\Phi$ is the vector of relative weights. Therefore, the full likelihood can be written as:
\begin{equation}
    \mathcal{L}_{\text{NN}}(\bm{d}_{PS} | \bm{\theta}) = \sum_{i=1}^K \phi_i(\bm{\theta}) \cdot \mathcal{N}(\bm{d}_{PS}| \bm{\mu}_{\text{NN}, i}(\bm{\theta}), \Sigma_{\text{NN}, i}(\bm{\theta})) \, ,
\end{equation}
where $\sum_i \phi_i(\bm{\theta}) = 1$. Such a parametrization will be able to catch both higher-order moments and possible multi-modalities in the likelihood.  Here we use $K=3$ Gaussians, as it provides the lowest validation loss.

We also use CMAF density estimation (\citealt{Papamakarios_2017, Papamakarios_2018}). A CMAF is parametrized through a series of linear transformations of normal random variables. The idea is to start from a unit-variance, zero-mean normal distribution $p_{\bm{u}}(\bm{u})$ and through the transform $\bm{x} = T(\bm{u})$ find the estimate of some final distribution $p_{\bm{x}}(\bm{x})$. If the transformation $T$ is invertible and differentiable, one can write the final distribution as
\begin{align}
    p_{\bm{x}}(\bm{x}) &= p_{\bm{u}}(\bm{u}) \, \left| \det J_T (\bm{u})\right|^{-1} \nonumber \\
    &= p_{\bm{u}}(T^{-1}(\bm{u})) \, \left| \det J_{T^{-1}} (\bm{x})\right| \, ,
\end{align}
where $J_T$ and $J_{T^{-1}}$ are Jacobians of $T$ and its inverse, respectively. For larger expressivity, one can stack several transformations so that $T = T_n \circ T_{n-1} \circ \ldots \circ T_1$. This will reflect in the above equation as the multiplication of $n$ Jacobian determinants. The parameters of the network are then parameters of the transformations, trained in the same way as before - by minimizing KL divergence between transformed distribution and the real one. In our case, $\bm{x} = (\bm{d}_{PS}, \bm{\theta})$ represents the space of the PS and parameters. In order to construct a likelihood, one then fixes parameter dimensions and transforms PS dimensions only. For more details, see a recent review by \cite{Papamakarios_2019}.  Although CMAF NDEs offer more flexibility to capture complicated likelihoods, they are less stable to train compared to Gaussian mixture NDEs.  Indeed, we use Gaussian mixture as our reference NDE below, as it has the best validation loss (see results)  and is the least biased (see Section \ref{sec:SBC}). 

For all NDEs except CMAF, the underlying NNs are defined with $10$ fully-connected layers of $100$ neurons, LeakyReLU (e.g. \citealt{Bing2015}) activation function (except for the last layer), followed by a layer defining the likelihood parameters. The CMAF NDE is of the same shape, using the same activation function, where the sequence of layers parametrizes the chain of transformations. Our NNs use the Adam optimizer \citep{Kingma2014}, learning rate reduction and early stopping. All NDE likelihoods and training procedures are made public as a part of \texttt{21cmLikelihoods}\footnote{\url{https://github.com/dprelogo/21cmLikelihoods}} package.

Finally, it is important to test that the likelihood density estimation has trained well, without introducing biases and being under/over confident.  One can use a mock observation and compare the resulting posteriors obtained from different likelihood estimators.  If the mock observation is not consistent with the recovered posterior, then the likelihood estimator is biased (Section \ref{sec:fiducial}).  If this is repeated for many mock observations, sampled from the joint distribution $P(\bm{\theta}, \bm{d}_{PS})$, the average over all posteriors should reproduce the prior.  This method (described in more details in Section \ref{sec:SBC}) is called simulation-based calibration (SBC; \citealt{Talts_2018}).  SBC provides a rigorous, quantitative test of the accuracy of likelihood estimation.  Unfortunately, it requires averaging over thousands of posteriors and is therefore only computationally tractable for amortized inference like SBI.

\section{Results} \label{ch:results}

In summary, we compare the performance of the following likelihood choices for the spherically-averaged 21-cm power spectrum:
\begin{enumerate}
\item {\bf CLASSIC fixed var} - assumes a Gaussian likelihood, with the mean estimated from a single (albeit well chosen) realization, and a diagonal covariance estimated only at $\bm{\theta}_{\rm fid}$. This is by far the most common choice in 21-cm inference. In fact, this is an improvement over the most common method of {\it randomly} picking a realization for the model. As mentioned above, we pick the one realization out of 1000 that is closest to the mean at the fiducial parameter choice.
\item {\bf CLASSIC fixed cov} - like {\bf CLASSIC fixed var}, but using a non-diagonal covariance.
\item {\bf NDE fixed var} - assumes a Gaussian likelihood, with the mean estimated by NDE, and a diagonal covariance estimated only at $\bm{\theta}_{\rm fid}$.
\item {\bf NDE fixed cov} - like  {\bf NDE fixed var}, but with a non-diagonal covariance.
\item {\bf NDE varying var} - assumes a Gaussian likelihood, with the mean and a diagonal covariance estimated throughout parameter space by NDE.
\item {\bf NDE varying cov} - like  {\bf NDE varying var}, but with the NDE fitting a non-diagonal covariance.
\item {\bf NDE CMAF} - a non-Gaussian likelihood estimator using CMAF NDE.
\item {\bf NDE Gauss mixture} - a non-Gaussian likelihood estimator using Gaussian mixture NDE. 
\end{enumerate}
For the first two \enquote{classic} likelihood choices, the simulator is called on-the-fly during inference (i.e. it is not amortized). The other likelihoods are used in SBI (amortized inference). Using validation loss and SBC rank statistics, (see Figures \ref{fig:training_losses} and \ref{fig:SBC}), we determine that the {\bf NDE Gauss mixture} is the most accurate of the likelihood estimators considered here. We therefore use it as a reference throughout the rest of this work. We summarize the assumptions for each likelihood choice in Table \ref{tab:likelihood_summary}.

\begin{table*}
\begin{tabular}{l|cccc}
                  & non-Gaussian & non-diagonal covariance & (co)variance is a function of $\bm{\theta}$ & mean by averaging over realizations  \\
\hline
CLASSIC fixed var & \xmark & \xmark & \xmark & \xmark ~ (single, well-chosen seed)  \\
CLASSIC fixed cov & \xmark & \cmark & \xmark & \xmark ~ (single, well-chosen seed) \\
NDE fixed var     & \xmark & \xmark & \xmark & \cmark  \\
NDE fixed cov     & \xmark & \cmark & \xmark & \cmark  \\
NDE varying var   & \xmark & \xmark & \cmark & \cmark  \\
NDE varying cov   & \xmark & \cmark & \cmark & \cmark  \\
NDE CMAF          & \cmark & N/A    & N/A    & N/A     \\
NDE Gauss mixture & \cmark & N/A    & N/A    & N/A                                                \end{tabular}
\caption{Properties of the likelihoods used in this study.  The CLASSIC likelihoods are used for non-amortized inference, in which the simulator is called at each likelihood evaluation.  The NDE likelihoods make use of a pre-computed (amortized) set of simulations to fit explicit (first four) or non-explicit (final two) likelihoods.  Based on the validation loss and SBC, we find {\it NDE Gauss mixture} to be the most accurate and stable to train.}
\label{tab:likelihood_summary}
\end{table*}

We compare the performances of the likelihood choices in two parts. In \S \ref{sec:fiducial} we show posteriors for the mock observation, and qualitatively describe their performances. In \S \ref{sec:generalization} we quantitatively test all NDE likelihoods using many mock observations across parameter space.

\subsection{Posteriors for the fiducial mock observation}
\label{sec:fiducial}
In order to recover the posteriors of the fiducial mock observation (see Figure \ref{fig:variance_likelihood}), we run Bayesian inference for the constructed and trained likelihoods. For the two classic choices, we use \texttt{21CMMC}\footnote{\url{https://github.com/21cmfast/21CMMC}} package, together with the \texttt{MultiNest}\footnote{\url{https://github.com/JohannesBuchner/PyMultiNest}} sampler \citep{Buchner_2014}. For the NDEs, we use \texttt{UltraNest}\footnote{\url{https://github.com/JohannesBuchner/UltraNest}} \citep{Buchner_2021}, with the MLFriends algorithm \citep{Buchner_2016, Buchner_2019}. For simple posteriors like ours the two codes perform the same; however, the computational cost of SBI is lower with \texttt{UltraNest}'s vectorized likelihood evaluation.  For our default convergence criteria that require $\sim$40k--90k likelihood evaluations, the non-amortized (classic) inference with \texttt{21CMMC} takes $\sim 10^5$ core hours, while the amortized (SBI) inference with \texttt{UltraNest} takes $\sim 0.1$ core hours.  We return to the relative computational costs in more detail below.

\begin{figure*}
\begin{minipage}[t]{0.47\linewidth}
    \centering
    \includegraphics[width=.99\linewidth]{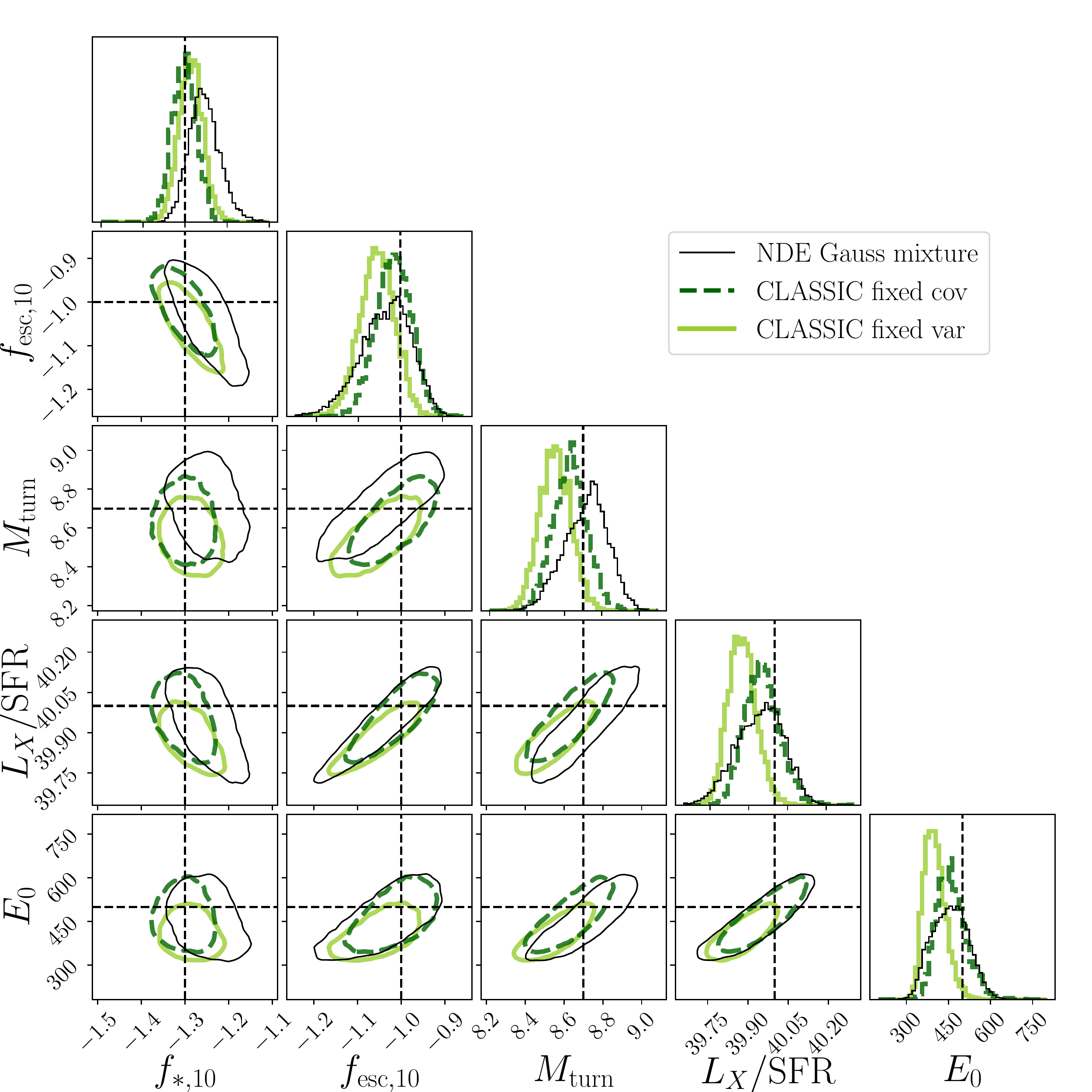}
    \vspace{4ex}
    \caption{Posterior of the fiducial mock observation recovered from the \textit{NDE Gauss mixture} likelihood, compared with \textit{CLASSIC fixed cov} and \textit{CLASSIC fixed var}. Individual figures show 2D or 1D marginalizations of the posterior. Contours label 95\% confidence intervals, and black-dashed lines true parameter values. }
    \label{fig:fiducial_posterior_1}
\end{minipage}
\hspace{0.04\linewidth}
\begin{minipage}[t]{0.47\linewidth}
    \centering
    \includegraphics[width=.99\linewidth]{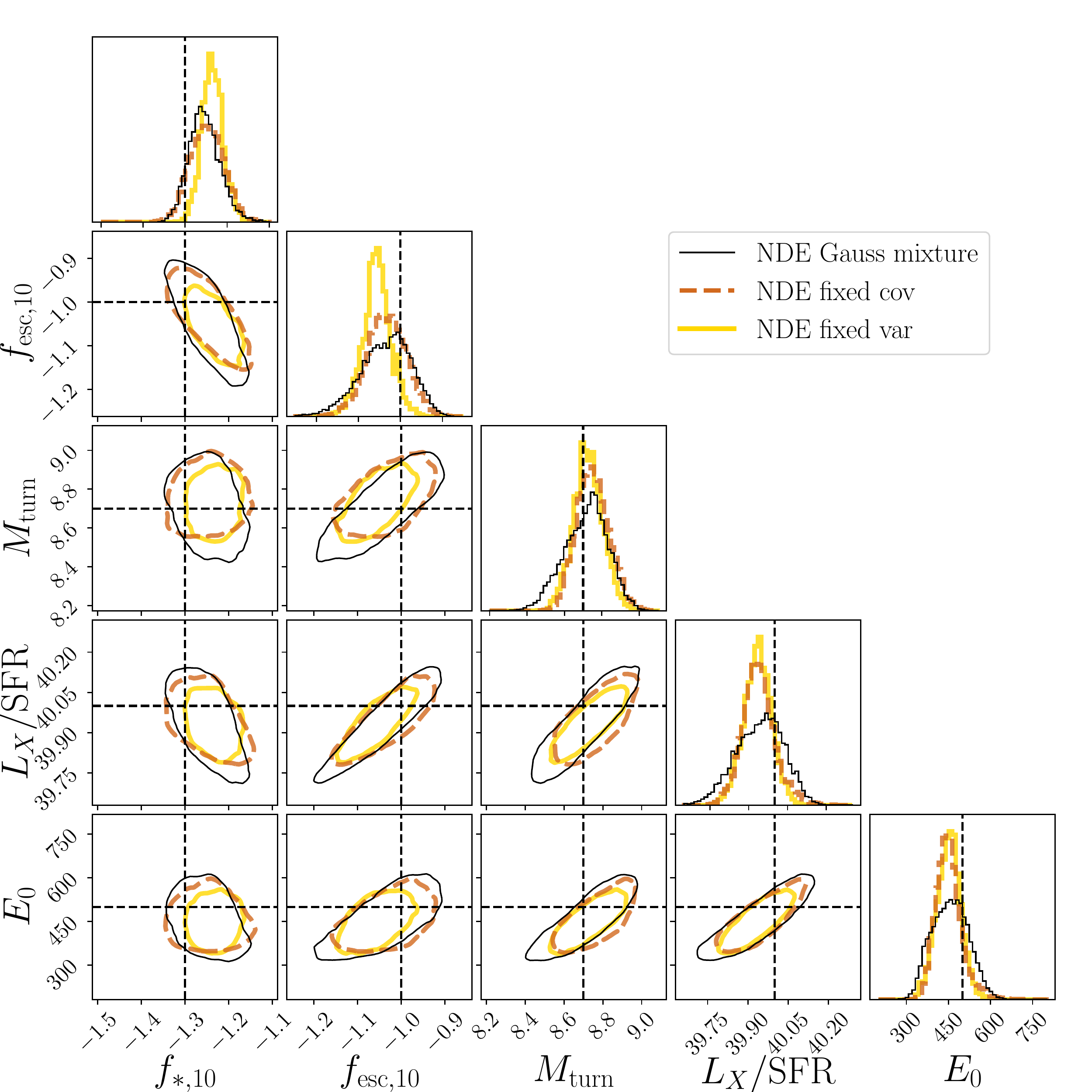}
    \vspace{4ex}
    \caption{Like Figure \ref{fig:fiducial_posterior_1}, but here we compare the posteriors from \textit{NDE fixed cov} and \textit{NDE fixed var}.}
    \label{fig:fiducial_posterior_2}
\end{minipage}
\\
\begin{minipage}[t]{0.47\linewidth}
    \centering
    \includegraphics[width=.99\linewidth]{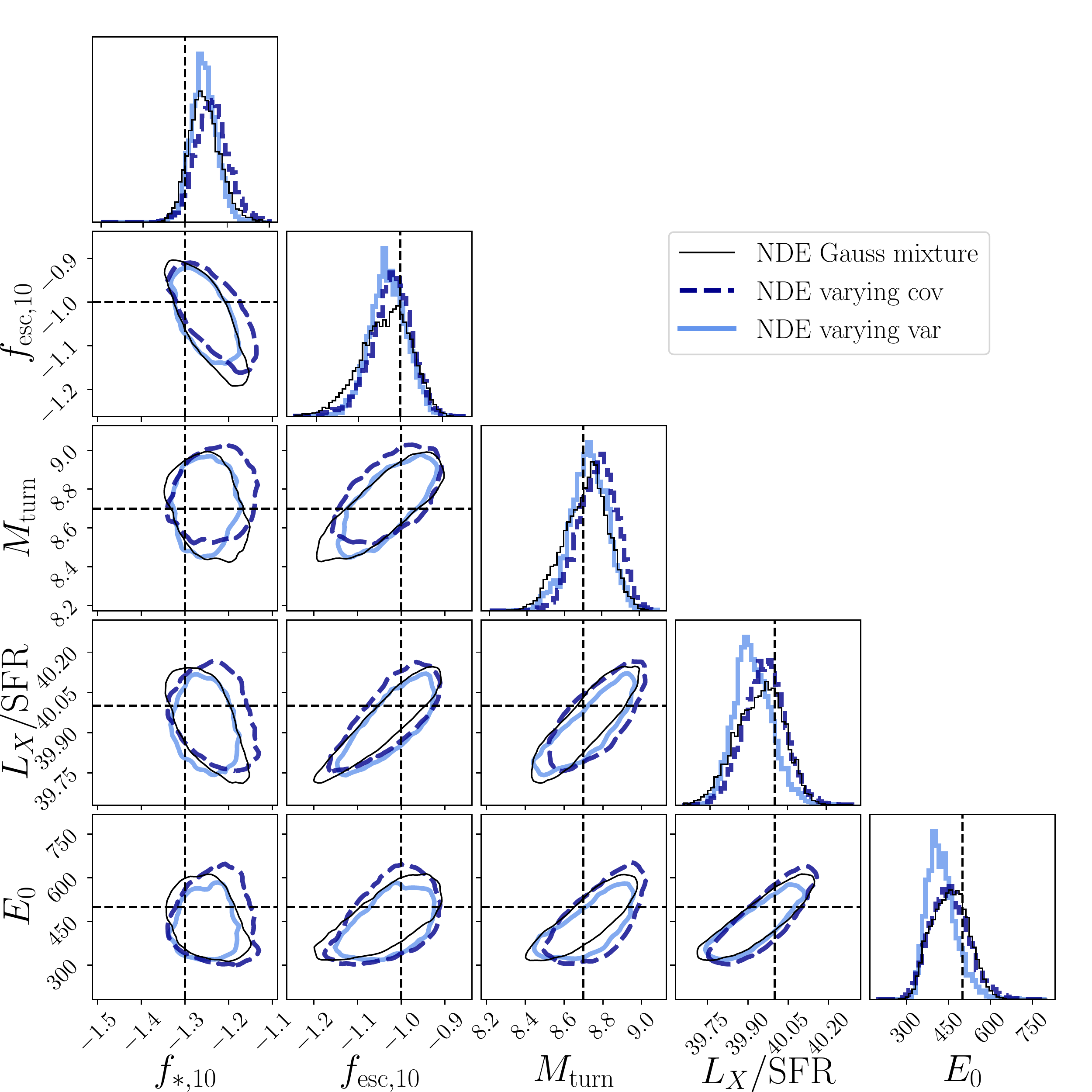}
    \vspace{4ex}
    \caption{Like Figure \ref{fig:fiducial_posterior_1}, but here we compare the posteriors from \textit{NDE varying cov} and \textit{NDE varying var}.}
    \label{fig:fiducial_posterior_3}
\end{minipage}
\hspace{0.04\linewidth}
\begin{minipage}[t]{0.47\linewidth}
    \centering
    \includegraphics[width=.99\linewidth]{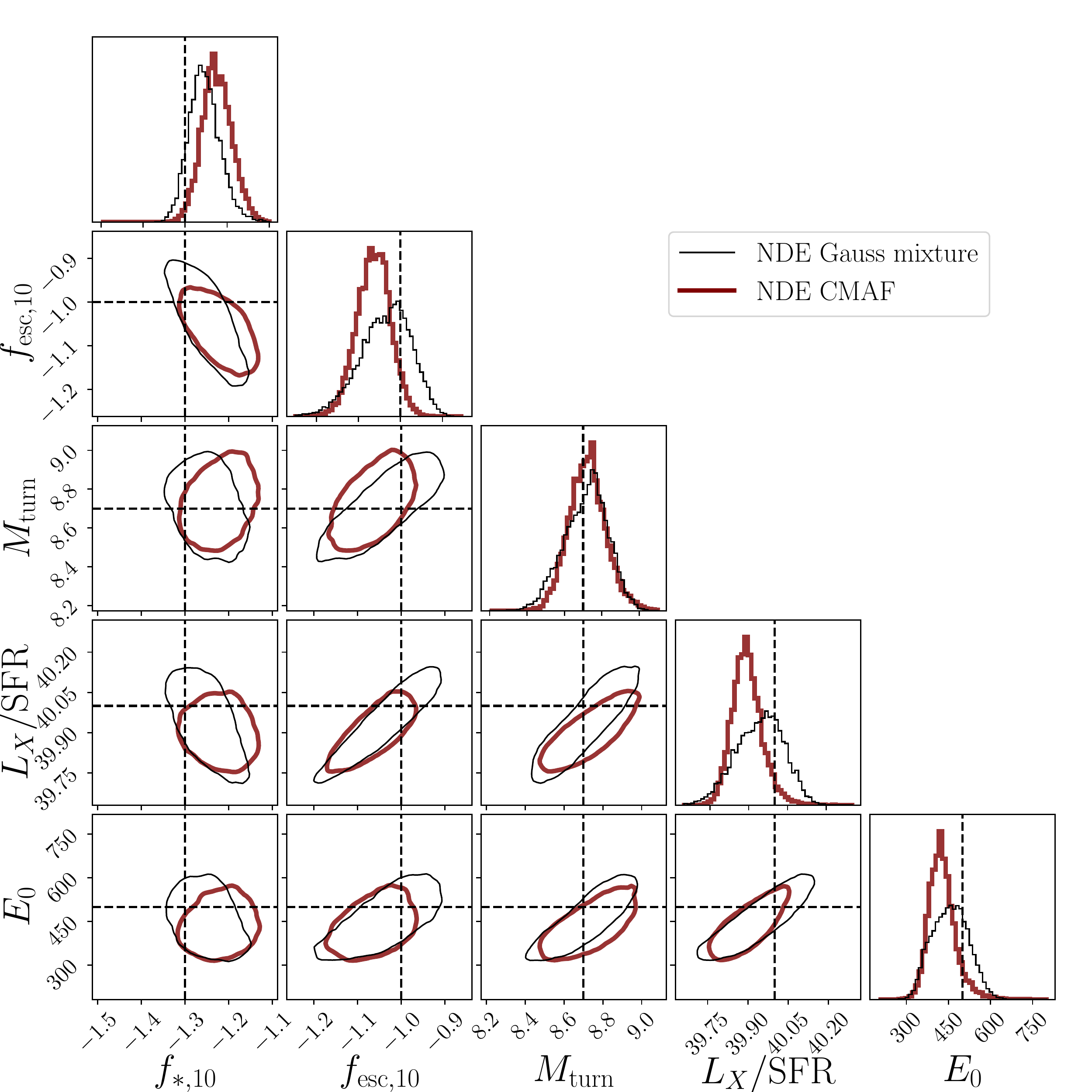}
    \vspace{4ex}
    \caption{Like Figure \ref{fig:fiducial_posterior_1}, but here comparing the two non-Gaussian likelihoods, \textit{NDE Gauss mixture} and  \textit{NDE CMAF}.}
    \label{fig:fiducial_posterior_4}
\end{minipage}
\end{figure*}

In Figure \ref{fig:fiducial_posterior_1}, we show 1D and 2D marginal posteriors for the two classic likelihoods in green, together with \textit{NDE Gauss mixture} as a reference in black. Straight lines denote the fiducial parameter values and 95\% C.I. are demarcated in the 2D panels. Comparing \textit{CLASSIC fixed var} with \textit{CLASSIC fixed cov}, we can clearly see that ignoring the non-diagonal covariance terms results in a higher bias and a more over-confident posterior.  This is in contrast with \citet{Zhao2022b} who find that for their two parameter model, the NDE likelihood results in a {\it tighter} posterior compared with a Gaussian likelihood. This difference could be due to the specific choice of model seed and/or the calculation of the variance (computed via Monte Carlo vs. assumed to be Poisson).  Regardless, we agree with their main conclusions that just using a diagonal Gaussian variance can bias inference.  Our true parameter values are in some cases inconsistent with the \textit{CLASSIC fixed var} posterior, {\it the most popular choice for 21-cm inference}, at greater than 95\% C.I.  Note that we confirm that our choice of data seed is not a rare outlier (see Fig. \ref{fig:variance_likelihood}), and so the truth should be inside the 68\% C.I.
This bias is especially obvious for the X-ray heating parameters, $E_0$ and $L_X/{\rm SFR}$.
Including a full covariance, still only evaluated at $\bm{\theta}_{\rm fid}$, gives a more accurate recovery.  Indeed, the true values of the fiducial mock are always consistent with the recovered \textit{CLASSIC fixed cov} posterior.  Because it only requires pre-computing a few hundred additional realizations of the fiducial model, the 
\textit{CLASSIC fixed cov} likelihood provides an easy way of increasing the precision of classical inference.  However, even though it is a significant improvement over  \textit{CLASSIC fixed var}, we see that \textit{CLASSIC fixed cov} is still too narrow compared with our reference, non-Gaussian posterior in black.

In Figure \ref{fig:fiducial_posterior_2}, we show results for \textit{NDE fixed cov} in brown and \textit{NDE fixed var} in yellow, with \textit{NDE Gauss mixture} for reference as before. Comparing to the classic likelihood results from Figure \ref{fig:fiducial_posterior_1}, we can see that the bias and constraints from \textit{NDE fixed var} are consistent with \textit{CLASSIC fixed var}, while those from \textit{NDE fixed cov} are consistent with \textit{NDE fixed var}.
There is a difference in the direction of the bias, which is expected as the mean in the two cases is estimated by different methods. 
We therefore conclude that using NDE to find the mean PS does not provide a significant improvement over using a single realization as the mean, {\it provided that the realization is carefully chosen to be the one closest to the mean at the fiducial parameter value.}  This last caveat is important, as most classic inferences {\it randomly} choose an initial condition seed to be used in inference.

In Figure \ref{fig:fiducial_posterior_3} we show posteriors from \textit{NDE varying var} and \textit{NDE varying cov}, which allow the (co)variance to be a function of $\bm{\theta}$. In comparison to the previous results for which the (co)variance is only evaluated at a single point, we see a notable reduction in bias.  In fact, \textit{NDE varying cov} results in a very similar posterior as our reference \textit{NDE Gauss mixture}.   This suggests that higher-order moments (beyond the 2-pt covariance), do not contribute significantly to the 21-cm PS likelihood.  As we demonstrate in the next section, \textit{NDE varying cov} also generalizes well across our parameter space of mock observations, making it our second most accurate likelihood.

Finally, in Figure \ref{fig:fiducial_posterior_4}, we compare the posteriors from our two non-Gaussian likelihoods: \textit{NDE Gauss mixture} and \textit{NDE CMAF}. Surprisingly, our CMAF results in an over-confident and biased posterior.  The fact that the CMAF is more expressive (i.e. is able to adapt to many different distribution shapes) makes it more difficult to train,  requiring either a larger training set or sequential (active) training \citep{Papamakarios_2018, Alsing_2019}. Thankfully, the 21-cm PS likelihood is simple enough to be well fit by the more stable \textit{NDE Gauss mixture}, and does not require the additional flexibility of CMAF density estimation. 

\begin{figure*}
    \centering
    \includegraphics[width=.99\linewidth]{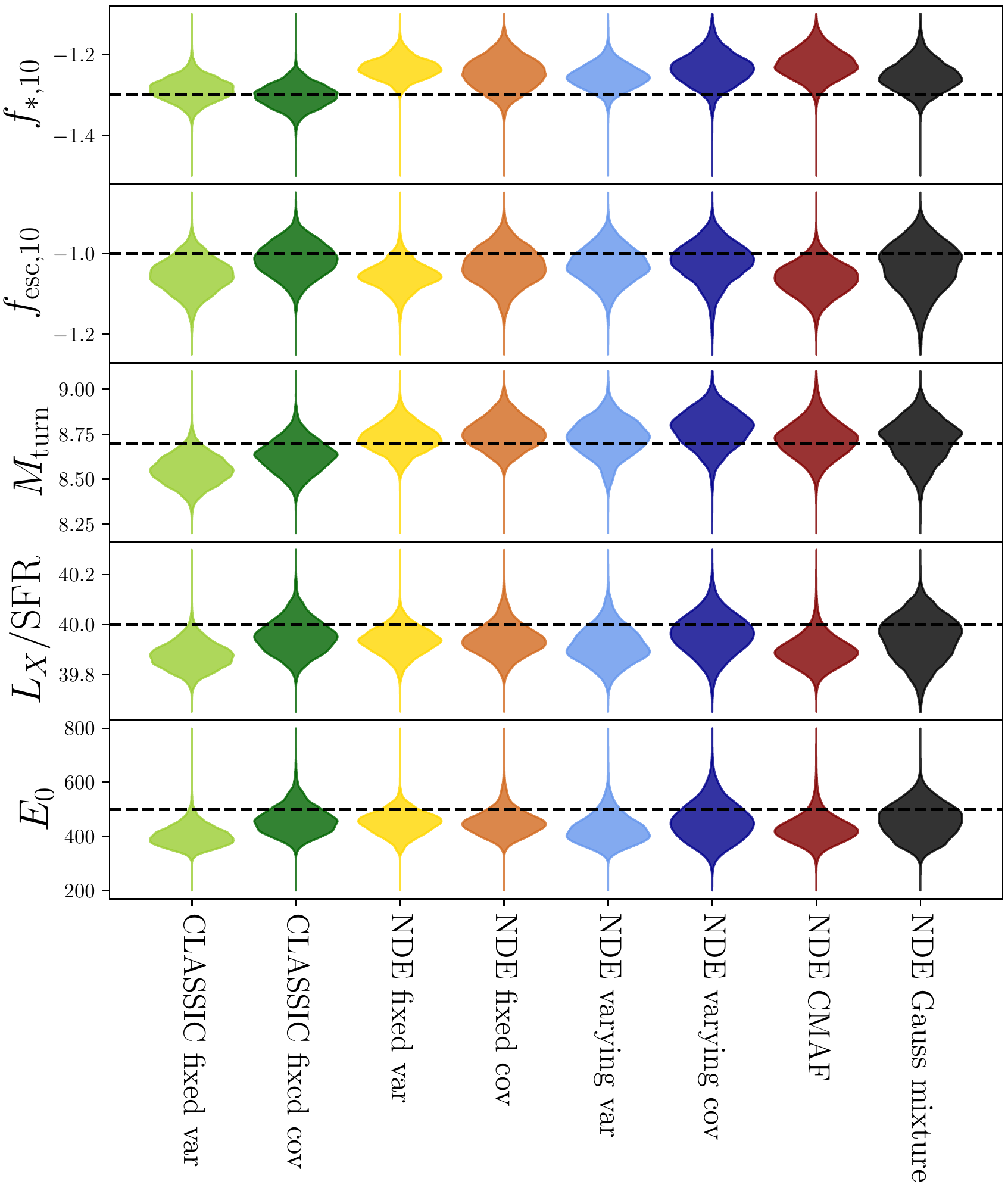}
    \caption{Summary of the previous posteriors, showing only 1D marginal PDFs for all likelihood choices. The fiducial (true) parameter values are denoted with horizontal, dashed lines. Most posteriors are overconfident.  The most accurate posterior estimate (see Section \ref{sec:SBC}), comes from the {\it NDE Gauss mixture}.}
    \label{fig:violin_plot}
\end{figure*}

We summarize our results in Figure \ref{fig:violin_plot}, by showing 1D distributions for all parameters (rows) and all likelihoods (columns). Black dashed lines denote the true parameter values used for the mock observation. For the \textit{NDE Gauss mixture} the true values are consistent with all of the 1D marginal PDFs. Although we cannot quantitatively judge the likelihood using only one posterior, we show in the following section that \textit{NDE Gauss mixture} performs the best in general.
\textit{NDE varying cov} follows it very closely, in both confidence and bias. \textit{NDE varying var} performs slightly worse.  The largest biases are seen for the fixed variance likelihoods.
These results confirm that it is important to account for the full covariance between wavemodes and redshifts, preferably at multiple points in parameter space.  Higher-order moments are not particularly important for characterizing the 21-cm PS likelihood.

\subsection{Do the likelihood estimators generalize well across parameter space?}
\label{sec:generalization}

In the previous section, we compared the posteriors resulting from our different likelihood estimators for a single mock observation.  In order to gain confidence in the ability of the likelihoods to generalize across parameter space, we must compare the posteriors for many mock observations generated at many different $\bm{\theta}$.  Here we quantify this using two metrics: (i) the validation loss; and (ii) rank distributions from SBC.  We discuss each in turn.

Unfortunately, it is impractical to generate thousands of classical, non-amortized inferences required for this assessment.  We therefore only use our SBI likelihoods in this section.  As was demonstrated in the previous section, {\it NDE fixed var} and {\it NDE fixed cov} can serve as proxies for  {\it CLASSIC fixed var} and {\it CLASSIC fixed cov}, respectively.

\subsubsection{Training and validation loss} \label{subsec:training}

Our loss function (Eq. \ref{eq:loss_function}) is an unnormalized estimate of the KL divergence between likelihood samples and the fitted distribution. For this reason, the validation loss is a measure of how well the fitted likelihood function generalizes over the whole parameter space.
Due to the missing normalization constant (Shannon entropy of the \enquote{real} distribution,  see Eq. \ref{eq:KL_divergence_explicit}), the loss function can be negative, while the KL divergence cannot.

\begin{figure}
    \centering
    \includegraphics[width=.99\linewidth]{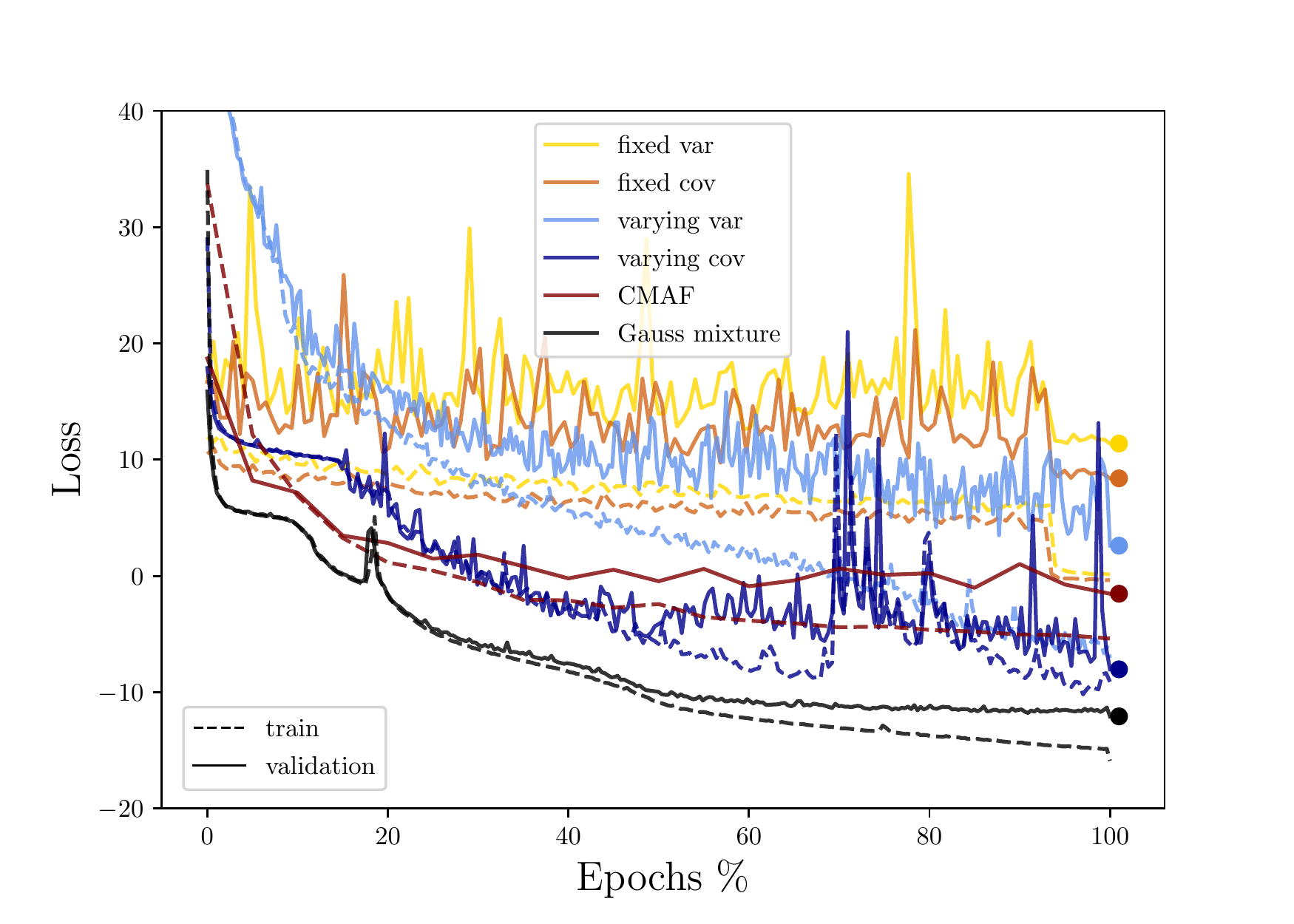}
    \caption{Training and validation losses for all NDEs. Loss on the $y$-axis corresponds to an unnormalized KL divergence and can be used as a relative measure of accuracy (see text for details).  On the $x$-axis we show the percent of the total number of epochs used for training (100\% corresponds to the minimum of the validation loss denoted with  dots).}
    \label{fig:training_losses}
\end{figure}

Figure \ref{fig:training_losses} shows the training and validation losses for all NDEs considered in this work, in dashed and solid lines, respectively. On the $x$-axis we show the percentage of the total number of epochs used for the final training (100\% corresponds to the minimum of the validation loss). The relative validation losses follow the same trends as noted in the previous section for the case of a single mock observation.  Namely, \textit{NDE Gauss mixture} has the lowest validation loss, confirming that it can accurately fit the likelihood across our parameter space.  \textit{NDE varying cov} is the second best, while \textit{NDE fixed var} is the worst.

It is useful to know just how large of a training set is needed to achieve the accuracy demonstrated by \textit{NDE Gauss mixture}.  Our default training set has 57k simulations.  This is roughly comparable to the average number of likelihood evaluations needed for Ultranest or Multinest to converge when doing inference\footnote{The exact number of likelihood evaluations depends not only on the mock observation, but also on the sampler settings such as the number of livepoints and the convergence criterion.  Our default setting uses 400 livepoints, requiring $\sim$90k likelihood evaluations.  We find that using 200 livepoints can give similar posteriors for our fiducial mock, requiring $\sim$40k likelihood evaluations.  Using a smaller number of livepoints results in noisy/biased posteriors.}.  The fact that these numbers are comparable means that amortized (running simulations before inference) and non-amortized (running simulations during inference) approaches to inference have comparable computational costs.  However, if our best performing \textit{NDE Gauss mixture} estimator can be trained with a smaller training set, then it would not only be more accurate but also faster than classical, non-amortized inference.

\begin{figure}
    \centering
    \includegraphics[width=.99\linewidth]{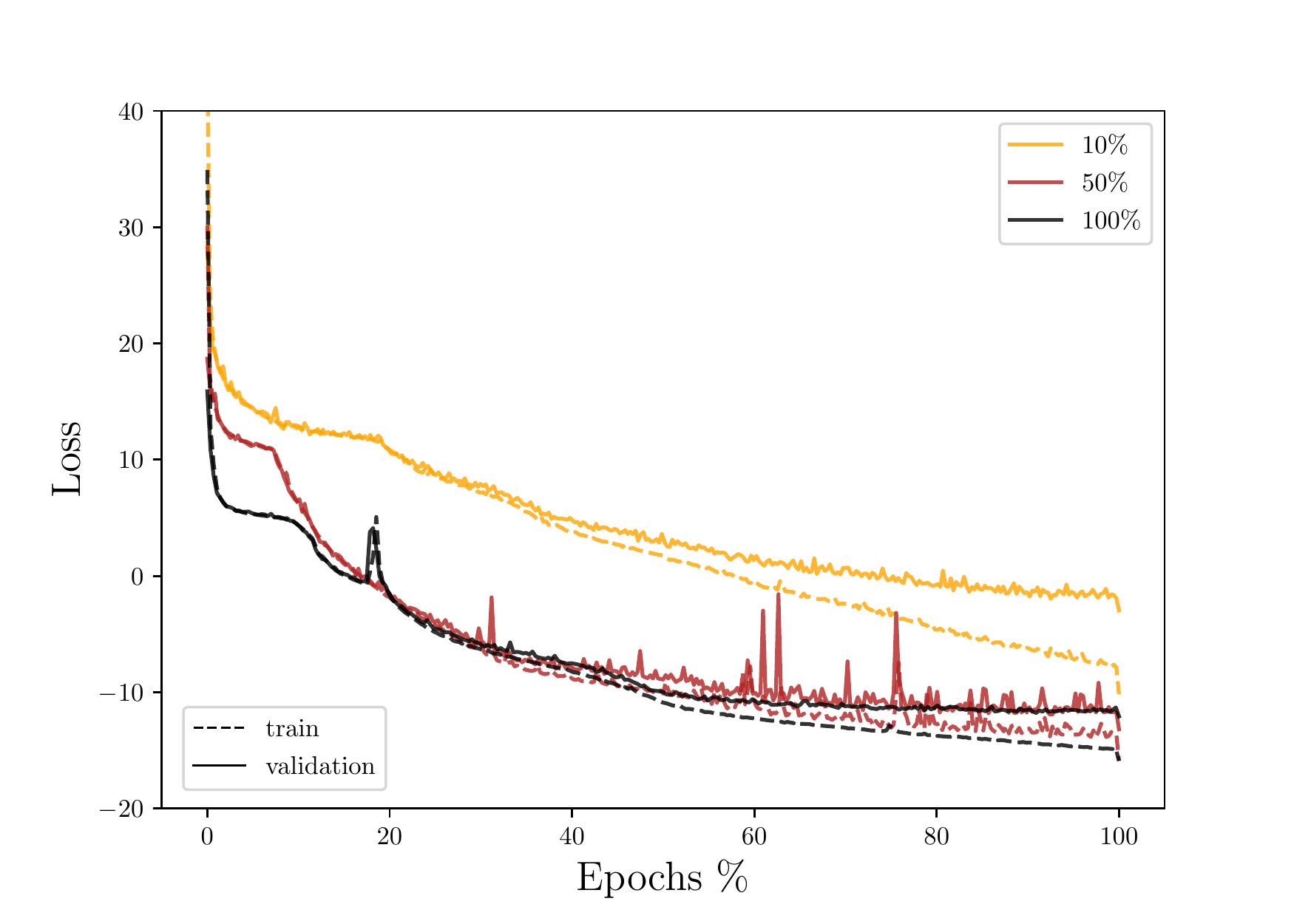}
    \caption{Training and validation losses for the {\it NDE Gauss mixture} model, for different training set sizes (denoted as a percentage of the full set, 57k, in the legend).  As in Figure \ref{fig:training_losses}, the loss on the $y$-axis is the unnormalized KL divergence and the $x$-axis shows the training percentage.}
    \label{fig:training_losses_gauss_mixture}
\end{figure}

To answer this question, we retrained \textit{NDE Gauss mixture} using only a subset of our training set: 50\% (28k simulations) and 10\% (5.7k simulations).\footnote{For visualization purposes, we keep the validation set fixed (17k samples).  This insures a smooth evolution of the validation loss curve; however we confirm that such a large validation set is not needed. We tested that for our usage case, having $1\,000$ samples was more than enough for the simple purpose of detecting over-fitting and reducing the learning rate if needed.}  The training and validation losses are plotted in Figure \ref{fig:training_losses_gauss_mixture}. One can clearly see that reducing the training size by a factor of produces little change in the validation loss. Therefore, 20k - 30k simulations are sufficient for \textit{NDE Gauss mixture} to maintain accuracy. Reducing the training size by a factor of ten pushes the validation loss to roughly the level of \textit{NDE varying cov} (see Fig. \ref{fig:training_losses}). Since this is our second best model, good performance is obtained even with $\sim$ 6k simulations.  {\it We thus conclude that SBI is not only more accurate than classic, non-amortized inference, but is faster by factors of $\gtrsim$3--10.}

The validation loss discussed in this section quantify the quality of the likelihood fits,  not the performance of the whole Bayesian inference framework. In the following section we  test the accuracy of the recovered posteriors and thus the robustness of our likelihood functions.

\subsubsection{Simulation-Based Calibration}
\label{sec:SBC}

In section \ref{sec:fiducial} we showed posteriors obtained from a single mock observation.  We compared the performances of different likelihood choices by noting if the true values from the mock were consistent with the recovered posteriors.  Here we extend this comparison over many different mock observations, in a procedure known as simulation-based calibration (SBC).

The core of the method is based on the following steps:
\begin{enumerate}
    \item sample parameters from prior,   $\widetilde{\bm{\theta}} \sim \pi(\bm{\theta})$
    \item sample the likelihood with a simulator,  $\widetilde{\bm{d}} \sim \mathcal{L}(\bm{d} | \widetilde{\bm{\theta}}) $
    \item calculate the posterior using the sample as a mock observation,  $\widetilde{\bm{d}} \rightarrow P(\bm{\theta} | \widetilde{\bm{d}}) $
\end{enumerate}
Here the samples are labeled with $\sim$. The first two steps are identical to SBI, with the final step adding the calculation of the posterior corresponding to that mock observation.
If we repeat these steps many times and average the posteriors for many $(\widetilde{\bm{\theta}}, \widetilde{\bm{d}})$, it follows that:
\begin{align}
    \pi(\bm{\theta}) &= \int P(\bm{\theta} | \widetilde{\bm{d}}) \cdot \mathcal{L}(\widetilde{\bm{d}} | \widetilde{\bm{\theta}}) \pi(\widetilde{\bm{\theta}}) \, \mathrm{d}\widetilde{\bm{\theta}} \, \mathrm{d}\widetilde{\bm{d}} \nonumber\\
    &\approx \frac{1}{N} \sum_{i = 1}^n P(\bm{\theta} | \widetilde{\bm{d}_i}) \, , \label{eq:data_avg_posterior}
\end{align}
where the last equation is the Monte-Carlo estimate of the integral for samples drawn from $P(\bm{\theta}, \bm{d}) = \mathcal{L}(\bm{d} | \bm{\theta}) \pi(\bm{\theta})$. In other words, \textit{prior = data-averaged posterior}. If any step of the Bayesian inference pipeline went wrong, the equality will not hold. This is the basis of SBC \citep{Talts_2018}. In particular, if we sample $L$ points $(\bm{\theta}_1, \ldots \bm{\theta}_L)$ from the posterior $P(\bm{\theta} | \widetilde{\bm{d}})$, make a scalar function $f: \Theta \rightarrow \mathbb{R}$ and define a rank function:
\begin{equation}
    r\left( f(\bm{\theta}_1), \ldots f(\bm{\theta}_L), f(\widetilde{\bm{\theta}})\right) = \sum_{l = 1}^L \mathbb{1}\left[f(\bm{\theta}_l) < f(\widetilde{\bm{\theta}})\right] \, , \label{eq:SBC_rank}
\end{equation}
a distribution of a rank $r$ should be uniform in $[0, L]$. Here the Boolean $\mathbb{1}$ equals $1$ if the condition is true, and $0$ otherwise. Therefore, the \textit{rank} of a particular posterior counts how many samples are below its true value. The \textit{Rank distribution} for rank $r$ counts how many posteriors have $r$ samples out of $L$ below their true values. This equation is a 1D representation of the data-averaged posterior in Eq. \ref{eq:data_avg_posterior}. By testing the uniformity in the rank $r$, one can see if posteriors are statistically under(over)-confident or biased. Over-confidence will make distribution convex ($\cup$-shaped), under-confidence concave ($\cap$-shaped) and any bias will tilt the distribution to one side. For instance, having a $\cap$-shaped profile in the case of under-confident posteriors can be intuitively understood as the following. Rank statistics measures the difference between the prior CDF and the data-averaged posterior CDF. If the two distributions are the same, the rank statistics are zero and flat.  However if the data-averaged posterior is wider, the low-rank points would move towards the middle ranks, as it is harder to obtain low ranks for a wide distribution. The opposite is true for the over-confident posteriors. For more details, see \citet{Talts_2018}.

We consider 5 scalar functions $f_i(\bm{\theta}) = \theta_i$, one for each astrophysical parameter.  This allows us to assess the quality of the 1D posteriors for each parameter separately. In order to calculate the rank statistics, we run \texttt{UltraNest} for all likelihoods and all samples in our test set of $\sim$ 7k simulations. \texttt{UltraNest} is the best choice for this task because of its: (i) vectorization of the likelihood calculation; (ii) robust convergence criteria; and (iii) non-correlated samples in the final posterior. The last point is crucial for SBC as correlated samples contaminate the rank statistics.  We calculate the ranks of the posterior samples via Eq. \ref{eq:SBC_rank}, for $L = 100$. 

\begin{figure*}
    \centering
    \includegraphics[width=.99\linewidth]{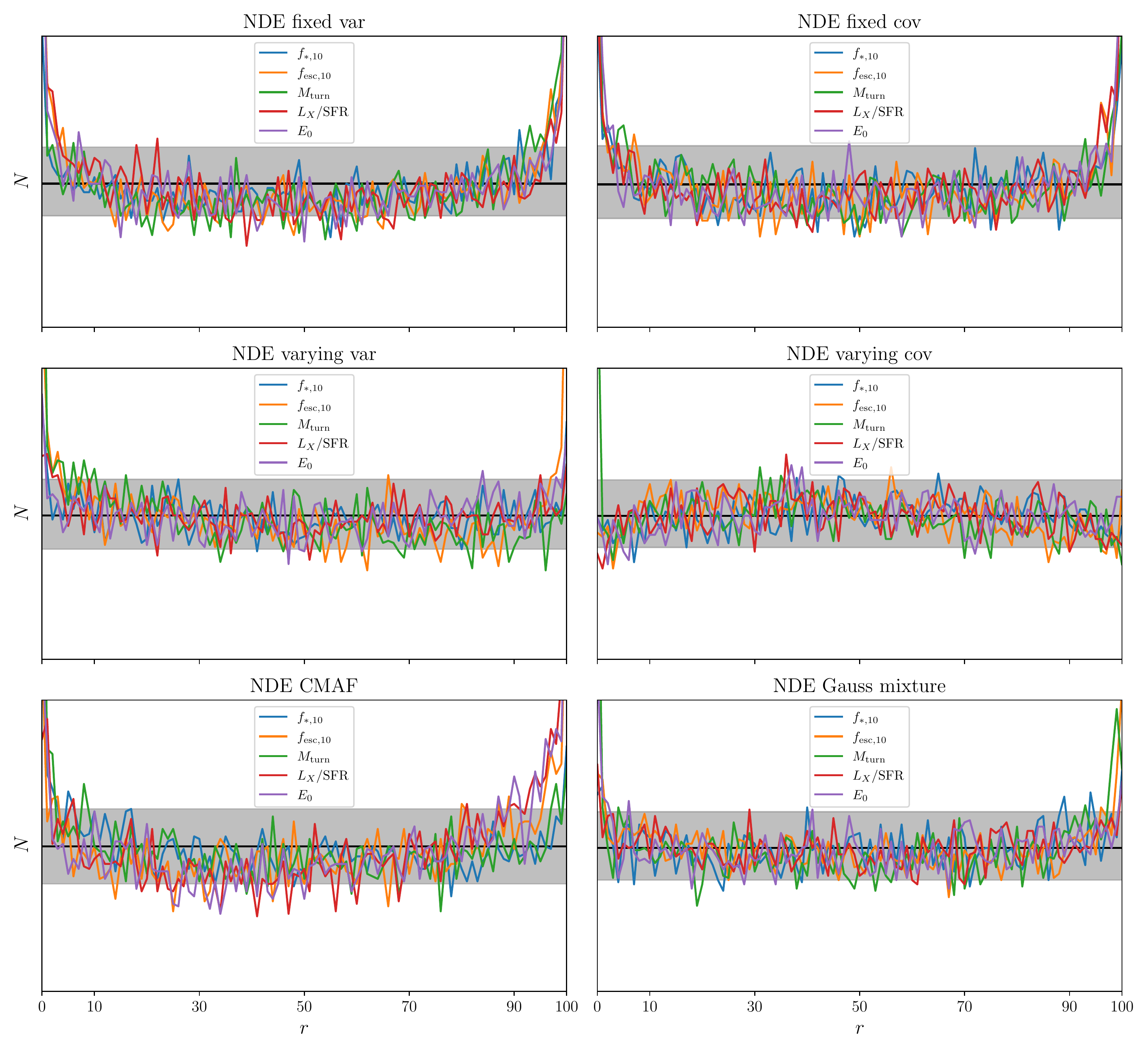}
    \caption{Posterior rank distributions from SBC computed from $\sim$7k models in the test set.  Each panel corresponds to a different NDE likelihood. The $x$-axis denotes the rank $r$ from 0 to $L = 100$. The $y$-axis shows the posterior number count $N$ of a particular rank. Black line and shaded area mark the mean and $95\%$ confidence interval of the binomial distribution, respectively.}
    \label{fig:SBC}
\end{figure*}

The results are shown in Figure \ref{fig:SBC}. Each rank statistic is expected to follow a binomial distribution - on the plot, the black line and gray area show its mean and $95\%$ interval, respectively. In general, the quality of the 1D posteriors is comparable within each likelihood choice (there is no obvious outlying astrophysical parameter).  For \textit{NDE fixed var} and \textit{NDE fixed cov} we can see a convex distribution, indicating that their posteriors are over-confident. In the case of \textit{NDE varying var} and \textit{NDE varying cov}, results look much better, with the curves staying in the $95\%$ region almost throughout the rank statistics.  Furthermore, \textit{NDE CMAF} seems to be fairly over-confident and slightly biased to the right, especially for recovery of $L_X / \text{SFR}$ and $E_0$.  

\textit{NDE Gauss mixture} performs the best across most of the rank statistic range, followed by {\it NDE varying cov}.  The former seems slightly overconfident while the latter slightly underconfident, but the distributions are generally consistent with the expected binomial 95\% interval in grey. This confirms the previous results that  \textit{NDE Gauss mixture} provides the most accurate likelihood.

\section{Conclusions} \label{ch:conclusions}

The spherically-averaged power spectrum is the most used summary statistic when performing inference from 21-cm tomography (using either current mock or upcoming data).  However, the likelihood functions commonly used in such inferences, contain several questionable assumptions.  Almost exclusively a Gaussian functional form is adopted, despite the fact the signal is non-Gaussian.  Even after assuming a Gaussian likelihood, further simplifications are made to make the likelihood evaluation computationally tractable.  These include: (i) estimating the mean from a single, random realization; (ii) assuming a diagonal covariance; (iii) computing the (co)variance only at a single point in parameter space.

Here we systematically test these assumptions with SBI: we train NDEs to fit the likelihood using a database of tens of thousands of simulated 21cm PS observations.  Our simulation pipeline consists of the following steps (c.f. Fig. \ref{fig:pipeline}): (i) sample the initial matter power spectrum and five galaxy parameters, computing a corresponding realization of the cosmic 21-cm lightcone with \cmfast{}; (ii) add a realization of the telescope noise corresponding to a 1000h integration with SKA1-Low; (iii) zero all Fourier modes lying within a foreground-dominated, horizon wedge; (iv) bin the lightcone in redshift intervals, computing the 1D 21-cm PS for each bin.  We make a database of 57k such realizations of the PS for training, including another 14k for validation and 7k for testing.  We compare the results of inference for various likelihood choices, validating the ability of our NDEs to generalize over parameter space using SBC.

Our main results are the following:
\begin{itemize}
\item The most common likelihood choice -- a Gaussian with a fixed variance evaluated only at a fiducial parameter -- results in an overconfident and biased posterior (see also \citealt{Zhao2022b}).  True values can be outside of the recovered 95\% C.I. when they should be inside 68\% C.I.
\item  Including the 2-point covariance between Fourier modes and redshifts dramatically improves inference.  If the covariance is fit as a function of the parameters, the resulting posteriors are almost identical to our best-performing, non-Gaussian NDE.  However, even pre-computing the covariance only at the fiducial parameter set gives decent results, making it a computationally-viable option for classic, non-amortized inference.
\item A single realization can be used in place of the mean PS across parameter space, provided that the realization is pre-selected to lie close to the mean at the fiducial parameter value.
\item Our most accurate posteriors result from a non-Gaussian likelihood, using a simple-to-train Gaussian mixture NDE.  More complex non-Gaussian NDEs like CMAFs are unneccessary and more difficult to train.
\item Our best-performing likelihood estimator can be trained on a database of only $\sim$20--30k simulations without any loss of accuracy.  Good performance (comparable to our second best estimator) can be achieved with only 6k simulations.  This is up to an order of magnitude smaller than the number of likelihood evaluations needed for inference.  Therefore creating a training set to fit the likelihood {\it before} running inference is more computationally efficient than calling a simulator {\it on-the-fly} with each likelihood evaluation during inference.
\end{itemize}

Although the choice of likelihood is unlikely to be important for current noisy 21-cm observations, we should have a high S/N detection of the cosmic 21-cm PS in the near future. This work highlights the power of SBI to provide accurate posteriors at a comparably low computational cost.

\section*{Acknowledgements}

We thank B. Greig, A. Liu, S. Murray, Y. Qin and B. Wandelt for helpful comments on a draft version of this manuscript.
A.M. acknowledges support from the Ministry of Universities and Research (MUR) through the PNRR project "Centro Nazionale di Ricerca in High Performance Computing, Big Data e Quantum Computing" and the PRO3 project "Data Science methods for Multi-Messenger Astrophysics and Cosmology".  We  gratefully acknowledge computational resources of the HPC center at SNS. This work used the Extreme Science and Engineering Discovery Environment (XSEDE), which is supported by National Science Foundation grant number ACI-1548562. Specifically, it used the Bridges-2 system, which is supported by NSF award number ACI-1928147, at the Pittsburgh Supercomputing Center (PSC).

\section*{Data Availability}

The data underlying this article will be shared on reasonable request to the corresponding author.


\bibliographystyle{mnras}
\bibliography{references} 

\begin{thebibliography}{}
\makeatletter
\relax
\def\mn@urlcharsother{\let\do\@makeother \do\$\do\&\do\#\do\^\do\_\do\%\do\~}
\def\mn@doi{\begingroup\mn@urlcharsother \@ifnextchar [ {\mn@doi@}
  {\mn@doi@[]}}
\def\mn@doi@[#1]#2{\def\@tempa{#1}\ifx\@tempa\@empty \href
  {http://dx.doi.org/#2} {doi:#2}\else \href {http://dx.doi.org/#2} {#1}\fi
  \endgroup}
\def\mn@eprint#1#2{\mn@eprint@#1:#2::\@nil}
\def\mn@eprint@arXiv#1{\href {http://arxiv.org/abs/#1} {{\tt arXiv:#1}}}
\def\mn@eprint@dblp#1{\href {http://dblp.uni-trier.de/rec/bibtex/#1.xml}
  {dblp:#1}}
\def\mn@eprint@#1:#2:#3:#4\@nil{\def\@tempa {#1}\def\@tempb {#2}\def\@tempc
  {#3}\ifx \@tempc \@empty \let \@tempc \@tempb \let \@tempb \@tempa \fi \ifx
  \@tempb \@empty \def\@tempb {arXiv}\fi \@ifundefined
  {mn@eprint@\@tempb}{\@tempb:\@tempc}{\expandafter \expandafter \csname
  mn@eprint@\@tempb\endcsname \expandafter{\@tempc}}}

\bibitem[\protect\citeauthoryear{{Abdurashidova} et~al.,}{{Abdurashidova}
  et~al.}{2022a}]{HERA_2022b}
{Abdurashidova} Z.,  et~al., 2022a, \mn@doi [\apj] {10.3847/1538-4357/ac2ffc},
  \href {https://ui.adsabs.harvard.edu/abs/2022ApJ...924...51A} {924, 51}

\bibitem[\protect\citeauthoryear{{Abdurashidova} et~al.,}{{Abdurashidova}
  et~al.}{2022b}]{HERA_2022a}
{Abdurashidova} Z.,  et~al., 2022b, \mn@doi [\apj] {10.3847/1538-4357/ac1c78},
  \href {https://ui.adsabs.harvard.edu/abs/2022ApJ...925..221A} {925, 221}

\bibitem[\protect\citeauthoryear{{Alsing}, {Wandelt}  \& {Feeney}}{{Alsing}
  et~al.}{2018}]{Alsing_2018}
{Alsing} J.,  {Wandelt} B.,   {Feeney} S.,  2018, \mn@doi [\mnras]
  {10.1093/mnras/sty819}, \href
  {https://ui.adsabs.harvard.edu/abs/2018MNRAS.477.2874A} {477, 2874}

\bibitem[\protect\citeauthoryear{{Alsing}, {Charnock}, {Feeney}  \&
  {Wandelt}}{{Alsing} et~al.}{2019}]{Alsing_2019}
{Alsing} J.,  {Charnock} T.,  {Feeney} S.,   {Wandelt} B.,  2019, \mn@doi
  [\mnras] {10.1093/mnras/stz1960}, \href
  {https://ui.adsabs.harvard.edu/abs/2019MNRAS.488.4440A} {488, 4440}

\bibitem[\protect\citeauthoryear{{Barkana} \& {Loeb}}{{Barkana} \&
  {Loeb}}{2004}]{barkana2004}
{Barkana} R.,  {Loeb} A.,  2004, \mn@doi [\apj] {10.1086/421079}, \href
  {https://ui.adsabs.harvard.edu/abs/2004ApJ...609..474B} {609, 474}

\bibitem[\protect\citeauthoryear{{Barkana} \& {Loeb}}{{Barkana} \&
  {Loeb}}{2005}]{Barkana2005}
{Barkana} R.,  {Loeb} A.,  2005, \mn@doi [\apj] {10.1086/429954}, \href
  {https://ui.adsabs.harvard.edu/abs/2005ApJ...626....1B} {626, 1}

\bibitem[\protect\citeauthoryear{{Bayer}, {Modi}  \& {Ferraro}}{{Bayer}
  et~al.}{2022}]{Bayer2022}
{Bayer} A.~E.,  {Modi} C.,   {Ferraro} S.,  2022, \mn@doi [arXiv e-prints]
  {10.48550/arXiv.2210.15649}, \href
  {https://ui.adsabs.harvard.edu/abs/2022arXiv221015649B} {p. arXiv:2210.15649}

\bibitem[\protect\citeauthoryear{{Behroozi}, {Wechsler}, {Hearin}  \&
  {Conroy}}{{Behroozi} et~al.}{2019}]{Behroozi_2019}
{Behroozi} P.,  {Wechsler} R.~H.,  {Hearin} A.~P.,   {Conroy} C.,  2019,
  \mn@doi [\mnras] {10.1093/mnras/stz1182}, \href
  {https://ui.adsabs.harvard.edu/abs/2019MNRAS.488.3143B} {488, 3143}

\bibitem[\protect\citeauthoryear{{Bouwens} et~al.,}{{Bouwens}
  et~al.}{2015a}]{Bouwens2015b}
{Bouwens} R.~J.,  et~al., 2015a, \mn@doi [\apj] {10.1088/0004-637X/803/1/34},
  \href {https://ui.adsabs.harvard.edu/abs/2015ApJ...803...34B} {803, 34}

\bibitem[\protect\citeauthoryear{{Bouwens}, {Illingworth}, {Oesch}, {Caruana},
  {Holwerda}, {Smit}  \& {Wilkins}}{{Bouwens} et~al.}{2015b}]{Bouwens_2015}
{Bouwens} R.~J.,  {Illingworth} G.~D.,  {Oesch} P.~A.,  {Caruana} J.,
  {Holwerda} B.,  {Smit} R.,   {Wilkins} S.,  2015b, \mn@doi [\apj]
  {10.1088/0004-637X/811/2/140}, \href
  {https://ui.adsabs.harvard.edu/abs/2015ApJ...811..140B} {811, 140}

\bibitem[\protect\citeauthoryear{{Bouwens}, {Oesch}, {Illingworth}, {Ellis}  \&
  {Stefanon}}{{Bouwens} et~al.}{2017}]{Bouwens2017}
{Bouwens} R.~J.,  {Oesch} P.~A.,  {Illingworth} G.~D.,  {Ellis} R.~S.,
  {Stefanon} M.,  2017, \mn@doi [\apj] {10.3847/1538-4357/aa70a4}, \href
  {https://ui.adsabs.harvard.edu/abs/2017ApJ...843..129B} {843, 129}

\bibitem[\protect\citeauthoryear{{Buchner}}{{Buchner}}{2016}]{Buchner_2016}
{Buchner} J.,  2016, \mn@doi [Statistics and Computing]
  {10.1007/s11222-014-9512-y}, \href
  {https://ui.adsabs.harvard.edu/abs/2016S&C....26..383B} {26, 383}

\bibitem[\protect\citeauthoryear{{Buchner}}{{Buchner}}{2019}]{Buchner_2019}
{Buchner} J.,  2019, \mn@doi [\pasp] {10.1088/1538-3873/aae7fc}, \href
  {https://ui.adsabs.harvard.edu/abs/2019PASP..131j8005B} {131, 108005}

\bibitem[\protect\citeauthoryear{{Buchner}}{{Buchner}}{2021}]{Buchner_2021}
{Buchner} J.,  2021, \mn@doi [The Journal of Open Source Software]
  {10.21105/joss.03001}, \href
  {https://ui.adsabs.harvard.edu/abs/2021JOSS....6.3001B} {6, 3001}

\bibitem[\protect\citeauthoryear{{Buchner} et~al.,}{{Buchner}
  et~al.}{2014}]{Buchner_2014}
{Buchner} J.,  et~al., 2014, \mn@doi [\aap] {10.1051/0004-6361/201322971},
  \href {https://ui.adsabs.harvard.edu/abs/2014A&A...564A.125B} {564, A125}

\bibitem[\protect\citeauthoryear{{Cole}, {Miller}, {Witte}, {Cai}, {Grootes},
  {Nattino}  \& {Weniger}}{{Cole} et~al.}{2022}]{Cole2022}
{Cole} A.,  {Miller} B.~K.,  {Witte} S.~J.,  {Cai} M.~X.,  {Grootes} M.~W.,
  {Nattino} F.,   {Weniger} C.,  2022, \mn@doi [\jcap]
  {10.1088/1475-7516/2022/09/004}, \href
  {https://ui.adsabs.harvard.edu/abs/2022JCAP...09..004C} {2022, 004}

\bibitem[\protect\citeauthoryear{{Cranmer}, {Brehmer}  \& {Louppe}}{{Cranmer}
  et~al.}{2020}]{cranmer_2020}
{Cranmer} K.,  {Brehmer} J.,   {Louppe} G.,  2020, \mn@doi [Proceedings of the
  National Academy of Science] {10.1073/pnas.1912789117}, \href
  {https://ui.adsabs.harvard.edu/abs/2020PNAS..11730055C} {117, 30055}

\bibitem[\protect\citeauthoryear{{Dai} \& {Seljak}}{{Dai} \&
  {Seljak}}{2022}]{Dai_2022}
{Dai} B.,  {Seljak} U.,  2022, \mn@doi [\mnras] {10.1093/mnras/stac2010}, \href
  {https://ui.adsabs.harvard.edu/abs/2022MNRAS.516.2363D} {516, 2363}

\bibitem[\protect\citeauthoryear{{Das}, {Mesinger}, {Pallottini}, {Ferrara}  \&
  {Wise}}{{Das} et~al.}{2017}]{Das_2017}
{Das} A.,  {Mesinger} A.,  {Pallottini} A.,  {Ferrara} A.,   {Wise} J.~H.,
  2017, \mn@doi [\mnras] {10.1093/mnras/stx943}, \href
  {https://ui.adsabs.harvard.edu/abs/2017MNRAS.469.1166D} {469, 1166}

\bibitem[\protect\citeauthoryear{{Fragos} et~al.,}{{Fragos}
  et~al.}{2013}]{Fragos2013}
{Fragos} T.,  et~al., 2013, \mn@doi [\apj] {10.1088/0004-637X/764/1/41}, \href
  {https://ui.adsabs.harvard.edu/abs/2013ApJ...764...41F} {764, 41}

\bibitem[\protect\citeauthoryear{{Furlanetto}, {Zaldarriaga}  \&
  {Hernquist}}{{Furlanetto} et~al.}{2004}]{furlanetto2004}
{Furlanetto} S.~R.,  {Zaldarriaga} M.,   {Hernquist} L.,  2004, \mn@doi [\apj]
  {10.1086/423025}, \href
  {https://ui.adsabs.harvard.edu/abs/2004ApJ...613....1F} {613, 1}

\bibitem[\protect\citeauthoryear{{Gazagnes}, {Koopmans}  \&
  {Wilkinson}}{{Gazagnes} et~al.}{2021}]{Gazagnes_2021}
{Gazagnes} S.,  {Koopmans} L. V.~E.,   {Wilkinson} M. H.~F.,  2021, \mn@doi
  [\mnras] {10.1093/mnras/stab107}, \href
  {https://ui.adsabs.harvard.edu/abs/2021MNRAS.502.1816G} {502, 1816}

\bibitem[\protect\citeauthoryear{{Ghara} et~al.,}{{Ghara}
  et~al.}{2020a}]{Ghara_2020}
{Ghara} R.,  et~al., 2020a, \mn@doi [\mnras] {10.1093/mnras/staa487}, \href
  {https://ui.adsabs.harvard.edu/abs/2020MNRAS.493.4728G} {493, 4728}

\bibitem[\protect\citeauthoryear{{Ghara} et~al.,}{{Ghara}
  et~al.}{2020b}]{emuPSG20}
{Ghara} R.,  et~al., 2020b, \mn@doi [\mnras] {10.1093/mnras/staa487}, \href
  {https://ui.adsabs.harvard.edu/abs/2020MNRAS.493.4728G} {493, 4728}

\bibitem[\protect\citeauthoryear{{Giri}, {Mellema}  \& {Jensen}}{{Giri}
  et~al.}{2020}]{tools21cm_JOSS}
{Giri} S.,  {Mellema} G.,   {Jensen} H.,  2020, \mn@doi [The Journal of Open
  Source Software] {10.21105/joss.02363}, \href
  {https://ui.adsabs.harvard.edu/abs/2020JOSS....5.2363G} {5, 2363}

\bibitem[\protect\citeauthoryear{{Giri}, {Schneider}, {Maion}  \&
  {Angulo}}{{Giri} et~al.}{2023}]{Giri2023}
{Giri} S.~K.,  {Schneider} A.,  {Maion} F.,   {Angulo} R.~E.,  2023, \mn@doi
  [\aap] {10.1051/0004-6361/202244986}, \href
  {https://ui.adsabs.harvard.edu/abs/2023A&A...669A...6G} {669, A6}

\bibitem[\protect\citeauthoryear{{Greig} \& {Mesinger}}{{Greig} \&
  {Mesinger}}{2015}]{Greig_2015}
{Greig} B.,  {Mesinger} A.,  2015, \mn@doi [\mnras] {10.1093/mnras/stv571},
  \href {https://ui.adsabs.harvard.edu/abs/2015MNRAS.449.4246G} {449, 4246}

\bibitem[\protect\citeauthoryear{{Greig} \& {Mesinger}}{{Greig} \&
  {Mesinger}}{2018}]{Greig_2018}
{Greig} B.,  {Mesinger} A.,  2018, \mn@doi [\mnras] {10.1093/mnras/sty796},
  \href {https://ui.adsabs.harvard.edu/abs/2018MNRAS.477.3217G} {477, 3217}

\bibitem[\protect\citeauthoryear{{Greig} et~al.,}{{Greig}
  et~al.}{2021}]{Greig_2021}
{Greig} B.,  et~al., 2021, \mn@doi [\mnras] {10.1093/mnras/staa3593}, \href
  {https://ui.adsabs.harvard.edu/abs/2021MNRAS.501....1G} {501, 1}

\bibitem[\protect\citeauthoryear{{Greig}, {Ting}  \& {Kaurov}}{{Greig}
  et~al.}{2022}]{Greig_2022}
{Greig} B.,  {Ting} Y.-S.,   {Kaurov} A.~A.,  2022, \mn@doi [\mnras]
  {10.1093/mnras/stac977}, \href
  {https://ui.adsabs.harvard.edu/abs/2022MNRAS.513.1719G} {513, 1719}

\bibitem[\protect\citeauthoryear{{Greig}, {Ting}  \& {Kaurov}}{{Greig}
  et~al.}{2023}]{Greig_2023}
{Greig} B.,  {Ting} Y.-S.,   {Kaurov} A.~A.,  2023, \mn@doi [\mnras]
  {10.1093/mnras/stac3822}, \href
  {https://ui.adsabs.harvard.edu/abs/2023MNRAS.519.5288G} {519, 5288}

\bibitem[\protect\citeauthoryear{{HERA Collaboration} et~al.,}{{HERA
  Collaboration} et~al.}{2023}]{HERA_2023}
{HERA Collaboration} et~al., 2023, \mn@doi [\apj] {10.3847/1538-4357/acaf50},
  \href {https://ui.adsabs.harvard.edu/abs/2023ApJ...945..124H} {945, 124}

\bibitem[\protect\citeauthoryear{{Jasche} \& {Kitaura}}{{Jasche} \&
  {Kitaura}}{2010}]{Jasche_2010}
{Jasche} J.,  {Kitaura} F.~S.,  2010, \mn@doi [\mnras]
  {10.1111/j.1365-2966.2010.16897.x}, \href
  {https://ui.adsabs.harvard.edu/abs/2010MNRAS.407...29J} {407, 29}

\bibitem[\protect\citeauthoryear{{Jasche} \& {Wandelt}}{{Jasche} \&
  {Wandelt}}{2012}]{Jasche_2012}
{Jasche} J.,  {Wandelt} B.~D.,  2012, \mn@doi [\mnras]
  {10.1111/j.1365-2966.2012.21423.x}, \href
  {https://ui.adsabs.harvard.edu/abs/2012MNRAS.425.1042J} {425, 1042}

\bibitem[\protect\citeauthoryear{{Jasche} \& {Wandelt}}{{Jasche} \&
  {Wandelt}}{2013}]{Jasche_2013}
{Jasche} J.,  {Wandelt} B.~D.,  2013, \mn@doi [\mnras] {10.1093/mnras/stt449},
  \href {https://ui.adsabs.harvard.edu/abs/2013MNRAS.432..894J} {432, 894}

\bibitem[\protect\citeauthoryear{{Jennings}, {Watkinson}, {Abdalla}  \&
  {McEwen}}{{Jennings} et~al.}{2019}]{emuPSJ19}
{Jennings} W.~D.,  {Watkinson} C.~A.,  {Abdalla} F.~B.,   {McEwen} J.~D.,
  2019, \mn@doi [\mnras] {10.1093/mnras/sty3168}, \href
  {https://ui.adsabs.harvard.edu/abs/2019MNRAS.483.2907J} {483, 2907}

\bibitem[\protect\citeauthoryear{{Jensen} et~al.,}{{Jensen}
  et~al.}{2013}]{Jensen2013}
{Jensen} H.,  et~al., 2013, \mn@doi [\mnras] {10.1093/mnras/stt1341}, \href
  {https://ui.adsabs.harvard.edu/abs/2013MNRAS.435..460J} {435, 460}

\bibitem[\protect\citeauthoryear{{Kern}, {Liu}, {Parsons}, {Mesinger}  \&
  {Greig}}{{Kern} et~al.}{2017a}]{Kern_2017}
{Kern} N.~S.,  {Liu} A.,  {Parsons} A.~R.,  {Mesinger} A.,   {Greig} B.,
  2017a, \mn@doi [\apj] {10.3847/1538-4357/aa8bb4}, \href
  {https://ui.adsabs.harvard.edu/abs/2017ApJ...848...23K} {848, 23}

\bibitem[\protect\citeauthoryear{{Kern}, {Liu}, {Parsons}, {Mesinger}  \&
  {Greig}}{{Kern} et~al.}{2017b}]{emuPSKern17}
{Kern} N.~S.,  {Liu} A.,  {Parsons} A.~R.,  {Mesinger} A.,   {Greig} B.,
  2017b, \mn@doi [\apj] {10.3847/1538-4357/aa8bb4}, \href
  {https://ui.adsabs.harvard.edu/abs/2017ApJ...848...23K} {848, 23}

\bibitem[\protect\citeauthoryear{{Kingma} \& {Ba}}{{Kingma} \&
  {Ba}}{2014}]{Kingma2014}
{Kingma} D.~P.,  {Ba} J.,  2014, \mn@doi [arXiv e-prints]
  {10.48550/arXiv.1412.6980}, \href
  {https://ui.adsabs.harvard.edu/abs/2014arXiv1412.6980K} {p. arXiv:1412.6980}

\bibitem[\protect\citeauthoryear{{Kitaura} \& {En{\ss}lin}}{{Kitaura} \&
  {En{\ss}lin}}{2008}]{Kitaura2008}
{Kitaura} F.~S.,  {En{\ss}lin} T.~A.,  2008, \mn@doi [\mnras]
  {10.1111/j.1365-2966.2008.13341.x}, \href
  {https://ui.adsabs.harvard.edu/abs/2008MNRAS.389..497K} {389, 497}

\bibitem[\protect\citeauthoryear{{Koopmans} et~al.,}{{Koopmans}
  et~al.}{2015}]{Koopmans2015}
{Koopmans} L.,  et~al., 2015, in Advancing Astrophysics with the Square
  Kilometre Array (AASKA14). p.~1 (\mn@eprint {arXiv} {1505.07568}),
  \mn@doi{10.22323/1.215.0001}

\bibitem[\protect\citeauthoryear{{Leclercq}, {Jasche}, {Lavaux}, {Wandelt}  \&
  {Percival}}{{Leclercq} et~al.}{2017}]{Leclercq_2017}
{Leclercq} F.,  {Jasche} J.,  {Lavaux} G.,  {Wandelt} B.,   {Percival} W.,
  2017, \mn@doi [\jcap] {10.1088/1475-7516/2017/06/049}, \href
  {https://ui.adsabs.harvard.edu/abs/2017JCAP...06..049L} {2017, 049}

\bibitem[\protect\citeauthoryear{{Lehmer} et~al.,}{{Lehmer}
  et~al.}{2016}]{Lehmer_2016}
{Lehmer} B.~D.,  et~al., 2016, \mn@doi [\apj] {10.3847/0004-637X/825/1/7},
  \href {https://ui.adsabs.harvard.edu/abs/2016ApJ...825....7L} {825, 7}

\bibitem[\protect\citeauthoryear{{Liu} \& {Shaw}}{{Liu} \&
  {Shaw}}{2020}]{Liu2020}
{Liu} A.,  {Shaw} J.~R.,  2020, \mn@doi [\pasp] {10.1088/1538-3873/ab5bfd},
  \href {https://ui.adsabs.harvard.edu/abs/2020PASP..132f2001L} {132, 062001}

\bibitem[\protect\citeauthoryear{{Liu}, {Parsons}  \& {Trott}}{{Liu}
  et~al.}{2014a}]{Liu2014a}
{Liu} A.,  {Parsons} A.~R.,   {Trott} C.~M.,  2014a, \mn@doi [\prd]
  {10.1103/PhysRevD.90.023018}, \href
  {https://ui.adsabs.harvard.edu/abs/2014PhRvD..90b3018L} {90, 023018}

\bibitem[\protect\citeauthoryear{{Liu}, {Parsons}  \& {Trott}}{{Liu}
  et~al.}{2014b}]{Liu2014b}
{Liu} A.,  {Parsons} A.~R.,   {Trott} C.~M.,  2014b, \mn@doi [\prd]
  {10.1103/PhysRevD.90.023019}, \href
  {https://ui.adsabs.harvard.edu/abs/2014PhRvD..90b3019L} {90, 023019}

\bibitem[\protect\citeauthoryear{{Loeb} \& {Zaldarriaga}}{{Loeb} \&
  {Zaldarriaga}}{2004}]{Loeb_2004}
{Loeb} A.,  {Zaldarriaga} M.,  2004, \mn@doi [\prl]
  {10.1103/PhysRevLett.92.211301}, \href
  {https://ui.adsabs.harvard.edu/abs/2004PhRvL..92u1301L} {92, 211301}

\bibitem[\protect\citeauthoryear{{Maity} \& {Choudhury}}{{Maity} \&
  {Choudhury}}{2023}]{Maity_2023}
{Maity} B.,  {Choudhury} T.~R.,  2023, \mn@doi [\mnras]
  {10.1093/mnras/stad791}, \href
  {https://ui.adsabs.harvard.edu/abs/2023MNRAS.tmp..740M} {}

\bibitem[\protect\citeauthoryear{{Mao}, {Shapiro}, {Mellema}, {Iliev}, {Koda}
  \& {Ahn}}{{Mao} et~al.}{2012}]{Mao_2012}
{Mao} Y.,  {Shapiro} P.~R.,  {Mellema} G.,  {Iliev} I.~T.,  {Koda} J.,   {Ahn}
  K.,  2012, \mn@doi [\mnras] {10.1111/j.1365-2966.2012.20471.x}, \href
  {https://ui.adsabs.harvard.edu/abs/2012MNRAS.422..926M} {422, 926}

\bibitem[\protect\citeauthoryear{{McAlpine} et~al.,}{{McAlpine}
  et~al.}{2022}]{McAlpine_2022}
{McAlpine} S.,  et~al., 2022, \mn@doi [\mnras] {10.1093/mnras/stac295}, \href
  {https://ui.adsabs.harvard.edu/abs/2022MNRAS.512.5823M} {512, 5823}

\bibitem[\protect\citeauthoryear{{McGreer}, {Mesinger}  \&
  {D'Odorico}}{{McGreer} et~al.}{2015}]{McGreer2015}
{McGreer} I.~D.,  {Mesinger} A.,   {D'Odorico} V.,  2015, \mn@doi [\mnras]
  {10.1093/mnras/stu2449}, \href
  {https://ui.adsabs.harvard.edu/abs/2015MNRAS.447..499M} {447, 499}

\bibitem[\protect\citeauthoryear{{McQuinn} \& {D'Aloisio}}{{McQuinn} \&
  {D'Aloisio}}{2018}]{McQuinn2018}
{McQuinn} M.,  {D'Aloisio} A.,  2018, \mn@doi [\jcap]
  {10.1088/1475-7516/2018/10/016}, \href
  {https://ui.adsabs.harvard.edu/abs/2018JCAP...10..016M} {2018, 016}

\bibitem[\protect\citeauthoryear{{McQuinn} \& {O'Leary}}{{McQuinn} \&
  {O'Leary}}{2012}]{McQuinn2012}
{McQuinn} M.,  {O'Leary} R.~M.,  2012, \mn@doi [\apj]
  {10.1088/0004-637X/760/1/3}, \href
  {https://ui.adsabs.harvard.edu/abs/2012ApJ...760....3M} {760, 3}

\bibitem[\protect\citeauthoryear{{Mellema} et~al.,}{{Mellema}
  et~al.}{2013}]{Mellema2013}
{Mellema} G.,  et~al., 2013, \mn@doi [Experimental Astronomy]
  {10.1007/s10686-013-9334-5}, \href
  {https://ui.adsabs.harvard.edu/abs/2013ExA....36..235M} {36, 235}

\bibitem[\protect\citeauthoryear{{Mertens} et~al.,}{{Mertens}
  et~al.}{2020}]{Mertens_2020}
{Mertens} F.~G.,  et~al., 2020, \mn@doi [\mnras] {10.1093/mnras/staa327}, \href
  {https://ui.adsabs.harvard.edu/abs/2020MNRAS.493.1662M} {493, 1662}

\bibitem[\protect\citeauthoryear{{Mesinger}}{{Mesinger}}{2020}]{Mesinger2020}
{Mesinger} A.,  2020, {The cosmic 21-cm revolution: charting the first billion
  years of our universe}, \mn@doi{10.1088/2514-3433/ab4a73.
}

\bibitem[\protect\citeauthoryear{{Mesinger} \& {Furlanetto}}{{Mesinger} \&
  {Furlanetto}}{2007}]{21cmFAST_Mesinger_07}
{Mesinger} A.,  {Furlanetto} S.,  2007, \mn@doi [\apj] {10.1086/521806}, \href
  {https://ui.adsabs.harvard.edu/abs/2007ApJ...669..663M} {669, 663}

\bibitem[\protect\citeauthoryear{{Mesinger}, {Furlanetto}  \& {Cen}}{{Mesinger}
  et~al.}{2011}]{21cmFAST_Mesinger11}
{Mesinger} A.,  {Furlanetto} S.,   {Cen} R.,  2011, \mn@doi [\mnras]
  {10.1111/j.1365-2966.2010.17731.x}, \href
  {https://ui.adsabs.harvard.edu/abs/2011MNRAS.411..955M} {411, 955}

\bibitem[\protect\citeauthoryear{{Mineo}, {Gilfanov}  \& {Sunyaev}}{{Mineo}
  et~al.}{2012}]{Mineo2012}
{Mineo} S.,  {Gilfanov} M.,   {Sunyaev} R.,  2012, \mn@doi [\mnras]
  {10.1111/j.1365-2966.2011.19862.x}, \href
  {https://ui.adsabs.harvard.edu/abs/2012MNRAS.419.2095M} {419, 2095}

\bibitem[\protect\citeauthoryear{{Mirocha} \& {Furlanetto}}{{Mirocha} \&
  {Furlanetto}}{2019}]{Mirocha_2019}
{Mirocha} J.,  {Furlanetto} S.~R.,  2019, \mn@doi [\mnras]
  {10.1093/mnras/sty3260}, \href
  {https://ui.adsabs.harvard.edu/abs/2019MNRAS.483.1980M} {483, 1980}

\bibitem[\protect\citeauthoryear{{Mondal}, {Bharadwaj}  \& {Majumdar}}{{Mondal}
  et~al.}{2017}]{Mondal2017}
{Mondal} R.,  {Bharadwaj} S.,   {Majumdar} S.,  2017, \mn@doi [\mnras]
  {10.1093/mnras/stw2599}, \href
  {https://ui.adsabs.harvard.edu/abs/2017MNRAS.464.2992M} {464, 2992}

\bibitem[\protect\citeauthoryear{{Mondal}, {Mellema}, {Murray}  \&
  {Greig}}{{Mondal} et~al.}{2022}]{emuPSM21}
{Mondal} R.,  {Mellema} G.,  {Murray} S.~G.,   {Greig} B.,  2022, \mn@doi
  [\mnras] {10.1093/mnrasl/slac053}, \href
  {https://ui.adsabs.harvard.edu/abs/2022MNRAS.514L..31M} {514, L31}

\bibitem[\protect\citeauthoryear{{Morales}, {Hazelton}, {Sullivan}  \&
  {Beardsley}}{{Morales} et~al.}{2012}]{Morales2012}
{Morales} M.~F.,  {Hazelton} B.,  {Sullivan} I.,   {Beardsley} A.,  2012,
  \mn@doi [\apj] {10.1088/0004-637X/752/2/137}, \href
  {https://ui.adsabs.harvard.edu/abs/2012ApJ...752..137M} {752, 137}

\bibitem[\protect\citeauthoryear{{Mu{\~n}oz}}{{Mu{\~n}oz}}{2023}]{Munoz2023}
{Mu{\~n}oz} J.~B.,  2023, \mn@doi [arXiv e-prints] {10.48550/arXiv.2302.08506},
  \href {https://ui.adsabs.harvard.edu/abs/2023arXiv230208506M} {p.
  arXiv:2302.08506}

\bibitem[\protect\citeauthoryear{{Murray} \& {Trott}}{{Murray} \&
  {Trott}}{2018}]{Murray2018}
{Murray} S.~G.,  {Trott} C.~M.,  2018, \mn@doi [\apj]
  {10.3847/1538-4357/aaebfa}, \href
  {https://ui.adsabs.harvard.edu/abs/2018ApJ...869...25M} {869, 25}

\bibitem[\protect\citeauthoryear{{Murray}, {Greig}, {Mesinger}, {Mu{\~n}oz},
  {Qin}, {Park}  \& {Watkinson}}{{Murray} et~al.}{2020}]{21cmFAST_JOSS}
{Murray} S.,  {Greig} B.,  {Mesinger} A.,  {Mu{\~n}oz} J.,  {Qin} Y.,  {Park}
  J.,   {Watkinson} C.,  2020, \mn@doi [The Journal of Open Source Software]
  {10.21105/joss.02582}, \href
  {https://ui.adsabs.harvard.edu/abs/2020JOSS....5.2582M} {5, 2582}

\bibitem[\protect\citeauthoryear{{Nasirudin}, {Murray}, {Trott}, {Greig},
  {Joseph}  \& {Power}}{{Nasirudin} et~al.}{2020}]{Nasirudin_2020}
{Nasirudin} A.,  {Murray} S.~G.,  {Trott} C.~M.,  {Greig} B.,  {Joseph} R.~C.,
   {Power} C.,  2020, \mn@doi [\apj] {10.3847/1538-4357/ab8003}, \href
  {https://ui.adsabs.harvard.edu/abs/2020ApJ...893..118N} {893, 118}

\bibitem[\protect\citeauthoryear{{O'Shea}, {Wise}, {Xu}  \& {Norman}}{{O'Shea}
  et~al.}{2015}]{OShea_2015}
{O'Shea} B.~W.,  {Wise} J.~H.,  {Xu} H.,   {Norman} M.~L.,  2015, \mn@doi
  [\apjl] {10.1088/2041-8205/807/1/L12}, \href
  {https://ui.adsabs.harvard.edu/abs/2015ApJ...807L..12O} {807, L12}

\bibitem[\protect\citeauthoryear{{Oesch}, {Bouwens}, {Illingworth}, {Labb{\'e}}
   \& {Stefanon}}{{Oesch} et~al.}{2018}]{Oesch2018}
{Oesch} P.~A.,  {Bouwens} R.~J.,  {Illingworth} G.~D.,  {Labb{\'e}} I.,
  {Stefanon} M.,  2018, \mn@doi [\apj] {10.3847/1538-4357/aab03f}, \href
  {https://ui.adsabs.harvard.edu/abs/2018ApJ...855..105O} {855, 105}

\bibitem[\protect\citeauthoryear{{Papamakarios}, {Pavlakou}  \&
  {Murray}}{{Papamakarios} et~al.}{2017}]{Papamakarios_2017}
{Papamakarios} G.,  {Pavlakou} T.,   {Murray} I.,  2017, \mn@doi [arXiv
  e-prints] {10.48550/arXiv.1705.07057}, \href
  {https://ui.adsabs.harvard.edu/abs/2017arXiv170507057P} {p. arXiv:1705.07057}

\bibitem[\protect\citeauthoryear{{Papamakarios}, {Sterratt}  \&
  {Murray}}{{Papamakarios} et~al.}{2018}]{Papamakarios_2018}
{Papamakarios} G.,  {Sterratt} D.~C.,   {Murray} I.,  2018, \mn@doi [arXiv
  e-prints] {10.48550/arXiv.1805.07226}, \href
  {https://ui.adsabs.harvard.edu/abs/2018arXiv180507226P} {p. arXiv:1805.07226}

\bibitem[\protect\citeauthoryear{{Papamakarios}, {Nalisnick}, {Jimenez
  Rezende}, {Mohamed}  \& {Lakshminarayanan}}{{Papamakarios}
  et~al.}{2019}]{Papamakarios_2019}
{Papamakarios} G.,  {Nalisnick} E.,  {Jimenez Rezende} D.,  {Mohamed} S.,
  {Lakshminarayanan} B.,  2019, \mn@doi [arXiv e-prints]
  {10.48550/arXiv.1912.02762}, \href
  {https://ui.adsabs.harvard.edu/abs/2019arXiv191202762P} {p. arXiv:1912.02762}

\bibitem[\protect\citeauthoryear{{Park}, {Mesinger}, {Greig}  \&
  {Gillet}}{{Park} et~al.}{2019}]{Park_2019}
{Park} J.,  {Mesinger} A.,  {Greig} B.,   {Gillet} N.,  2019, \mn@doi [\mnras]
  {10.1093/mnras/stz032}, \href
  {https://ui.adsabs.harvard.edu/abs/2019MNRAS.484..933P} {484, 933}

\bibitem[\protect\citeauthoryear{{Parsons} et~al.,}{{Parsons}
  et~al.}{2014}]{Parsons2014}
{Parsons} A.~R.,  et~al., 2014, \mn@doi [\apj] {10.1088/0004-637X/788/2/106},
  \href {https://ui.adsabs.harvard.edu/abs/2014ApJ...788..106P} {788, 106}

\bibitem[\protect\citeauthoryear{{Planck Collaboration} et~al.,}{{Planck
  Collaboration} et~al.}{2016}]{Planck_2016}
{Planck Collaboration} et~al., 2016, \mn@doi [\aap]
  {10.1051/0004-6361/201628897}, \href
  {https://ui.adsabs.harvard.edu/abs/2016A&A...596A.108P} {596, A108}

\bibitem[\protect\citeauthoryear{{Planck Collaboration} et~al.,}{{Planck
  Collaboration} et~al.}{2020}]{Planck_2018VI}
{Planck Collaboration} et~al., 2020, \mn@doi [\aap]
  {10.1051/0004-6361/201833910}, \href
  {https://ui.adsabs.harvard.edu/abs/2020A&A...641A...6P} {641, A6}

\bibitem[\protect\citeauthoryear{{Pober} et~al.,}{{Pober}
  et~al.}{2014}]{Pober2014}
{Pober} J.~C.,  et~al., 2014, \mn@doi [\apj] {10.1088/0004-637X/782/2/66},
  \href {https://ui.adsabs.harvard.edu/abs/2014ApJ...782...66P} {782, 66}

\bibitem[\protect\citeauthoryear{{Prelogovi{\'c}}, {Mesinger}, {Murray},
  {Fiameni}  \& {Gillet}}{{Prelogovi{\'c}} et~al.}{2022}]{prelogovic_22}
{Prelogovi{\'c}} D.,  {Mesinger} A.,  {Murray} S.,  {Fiameni} G.,   {Gillet}
  N.,  2022, \mn@doi [\mnras] {10.1093/mnras/stab3215}, \href
  {https://ui.adsabs.harvard.edu/abs/2022MNRAS.509.3852P} {509, 3852}

\bibitem[\protect\citeauthoryear{{Pritchard} \& {Furlanetto}}{{Pritchard} \&
  {Furlanetto}}{2007}]{Pritchard2007}
{Pritchard} J.~R.,  {Furlanetto} S.~R.,  2007, \mn@doi [\mnras]
  {10.1111/j.1365-2966.2007.11519.x}, \href
  {https://ui.adsabs.harvard.edu/abs/2007MNRAS.376.1680P} {376, 1680}

\bibitem[\protect\citeauthoryear{{Saxena}, {Cole}, {Gazagnes}, {Meerburg},
  {Weniger}  \& {Witte}}{{Saxena} et~al.}{2023}]{Saxena2023}
{Saxena} A.,  {Cole} A.,  {Gazagnes} S.,  {Meerburg} P.~D.,  {Weniger} C.,
  {Witte} S.~J.,  2023, \mn@doi [arXiv e-prints] {10.48550/arXiv.2303.07339},
  \href {https://ui.adsabs.harvard.edu/abs/2023arXiv230307339S} {p.
  arXiv:2303.07339}

\bibitem[\protect\citeauthoryear{Schmit \& Pritchard}{Schmit \&
  Pritchard}{2017}]{emuPSSP17}
Schmit C.~J.,  Pritchard J.~R.,  2017, \mn@doi [Monthly Notices of the Royal
  Astronomical Society] {10.1093/mnras/stx3292}, 475, 1213

\bibitem[\protect\citeauthoryear{{Schneider}, {Giri}  \& {Mirocha}}{{Schneider}
  et~al.}{2021}]{Schneider2021}
{Schneider} A.,  {Giri} S.~K.,   {Mirocha} J.,  2021, \mn@doi [\prd]
  {10.1103/PhysRevD.103.083025}, \href
  {https://ui.adsabs.harvard.edu/abs/2021PhRvD.103h3025S} {103, 083025}

\bibitem[\protect\citeauthoryear{{Scoccimarro}}{{Scoccimarro}}{1998}]{Scoccimarro_1998}
{Scoccimarro} R.,  1998, \mn@doi [\mnras] {10.1046/j.1365-8711.1998.01845.x},
  \href {https://ui.adsabs.harvard.edu/abs/1998MNRAS.299.1097S} {299, 1097}

\bibitem[\protect\citeauthoryear{{Shaw}, {Bharadwaj}  \& {Mondal}}{{Shaw}
  et~al.}{2019}]{Shaw2019}
{Shaw} A.~K.,  {Bharadwaj} S.,   {Mondal} R.,  2019, \mn@doi [\mnras]
  {10.1093/mnras/stz1561}, \href
  {https://ui.adsabs.harvard.edu/abs/2019MNRAS.487.4951S} {487, 4951}

\bibitem[\protect\citeauthoryear{{Shimabukuro} \& {Semelin}}{{Shimabukuro} \&
  {Semelin}}{2017}]{emuPSinvS17}
{Shimabukuro} H.,  {Semelin} B.,  2017, \mn@doi [\mnras]
  {10.1093/mnras/stx734}, \href
  {https://ui.adsabs.harvard.edu/abs/2017MNRAS.468.3869S} {468, 3869}

\bibitem[\protect\citeauthoryear{{Shimabukuro}, {Yoshiura}, {Takahashi},
  {Yokoyama}  \& {Ichiki}}{{Shimabukuro}
  et~al.}{2017}]{shimabukuro2017constraining}
{Shimabukuro} H.,  {Yoshiura} S.,  {Takahashi} K.,  {Yokoyama} S.,   {Ichiki}
  K.,  2017, \mn@doi [\mnras] {10.1093/mnras/stx530}, \href
  {https://ui.adsabs.harvard.edu/abs/2017MNRAS.468.1542S} {468, 1542}

\bibitem[\protect\citeauthoryear{{Sobacchi} \& {Mesinger}}{{Sobacchi} \&
  {Mesinger}}{2013a}]{Sobacchi2013}
{Sobacchi} E.,  {Mesinger} A.,  2013a, \mn@doi [\mnras]
  {10.1093/mnrasl/slt035}, \href
  {https://ui.adsabs.harvard.edu/abs/2013MNRAS.432L..51S} {432, L51}

\bibitem[\protect\citeauthoryear{{Sobacchi} \& {Mesinger}}{{Sobacchi} \&
  {Mesinger}}{2013b}]{Sobacchi2013b}
{Sobacchi} E.,  {Mesinger} A.,  2013b, \mn@doi [\mnras] {10.1093/mnras/stt693},
  \href {https://ui.adsabs.harvard.edu/abs/2013MNRAS.432.3340S} {432, 3340}

\bibitem[\protect\citeauthoryear{{Sobacchi} \& {Mesinger}}{{Sobacchi} \&
  {Mesinger}}{2014}]{sobacchi2014}
{Sobacchi} E.,  {Mesinger} A.,  2014, \mn@doi [\mnras] {10.1093/mnras/stu377},
  \href {https://ui.adsabs.harvard.edu/abs/2014MNRAS.440.1662S} {440, 1662}

\bibitem[\protect\citeauthoryear{{Talts}, {Betancourt}, {Simpson}, {Vehtari}
  \& {Gelman}}{{Talts} et~al.}{2018}]{Talts_2018}
{Talts} S.,  {Betancourt} M.,  {Simpson} D.,  {Vehtari} A.,   {Gelman} A.,
  2018, \mn@doi [arXiv e-prints] {10.48550/arXiv.1804.06788}, \href
  {https://ui.adsabs.harvard.edu/abs/2018arXiv180406788T} {p. arXiv:1804.06788}

\bibitem[\protect\citeauthoryear{{Trott}, {Wayth}  \& {Tingay}}{{Trott}
  et~al.}{2012}]{Trott2012}
{Trott} C.~M.,  {Wayth} R.~B.,   {Tingay} S.~J.,  2012, \mn@doi [\apj]
  {10.1088/0004-637X/757/1/101}, \href
  {https://ui.adsabs.harvard.edu/abs/2012ApJ...757..101T} {757, 101}

\bibitem[\protect\citeauthoryear{{Trott} et~al.,}{{Trott}
  et~al.}{2020}]{Trott_2020}
{Trott} C.~M.,  et~al., 2020, \mn@doi [\mnras] {10.1093/mnras/staa414}, \href
  {https://ui.adsabs.harvard.edu/abs/2020MNRAS.493.4711T} {493, 4711}

\bibitem[\protect\citeauthoryear{{Vedantham}, {Udaya Shankar}  \&
  {Subrahmanyan}}{{Vedantham} et~al.}{2012}]{Vedantham2012}
{Vedantham} H.,  {Udaya Shankar} N.,   {Subrahmanyan} R.,  2012, \mn@doi [\apj]
  {10.1088/0004-637X/745/2/176}, \href
  {https://ui.adsabs.harvard.edu/abs/2012ApJ...745..176V} {745, 176}

\bibitem[\protect\citeauthoryear{{Watkinson}, {Greig}  \&
  {Mesinger}}{{Watkinson} et~al.}{2022}]{Watkinson_2022}
{Watkinson} C.~A.,  {Greig} B.,   {Mesinger} A.,  2022, \mn@doi [\mnras]
  {10.1093/mnras/stab3706}, \href
  {https://ui.adsabs.harvard.edu/abs/2022MNRAS.510.3838W} {510, 3838}

\bibitem[\protect\citeauthoryear{{Xu}, {Wang}, {Chen}  \& {Li}}{{Xu}
  et~al.}{2015}]{Bing2015}
{Xu} B.,  {Wang} N.,  {Chen} T.,   {Li} M.,  2015, \mn@doi [arXiv e-prints]
  {10.48550/arXiv.1505.00853}, \href
  {https://ui.adsabs.harvard.edu/abs/2015arXiv150500853X} {p. arXiv:1505.00853}

\bibitem[\protect\citeauthoryear{{Xu}, {Wise}, {Norman}, {Ahn}  \&
  {O'Shea}}{{Xu} et~al.}{2016}]{Xu_2016}
{Xu} H.,  {Wise} J.~H.,  {Norman} M.~L.,  {Ahn} K.,   {O'Shea} B.~W.,  2016,
  \mn@doi [\apj] {10.3847/1538-4357/833/1/84}, \href
  {https://ui.adsabs.harvard.edu/abs/2016ApJ...833...84X} {833, 84}

\bibitem[\protect\citeauthoryear{{Zhao}, {Mao}, {Cheng}  \& {Wandelt}}{{Zhao}
  et~al.}{2022a}]{Zhao2022}
{Zhao} X.,  {Mao} Y.,  {Cheng} C.,   {Wandelt} B.~D.,  2022a, \mn@doi [\apj]
  {10.3847/1538-4357/ac457d}, \href
  {https://ui.adsabs.harvard.edu/abs/2022ApJ...926..151Z} {926, 151}

\bibitem[\protect\citeauthoryear{{Zhao}, {Mao}  \& {Wandelt}}{{Zhao}
  et~al.}{2022b}]{Zhao2022b}
{Zhao} X.,  {Mao} Y.,   {Wandelt} B.~D.,  2022b, \mn@doi [\apj]
  {10.3847/1538-4357/ac778e}, \href
  {https://ui.adsabs.harvard.edu/abs/2022ApJ...933..236Z} {933, 236}

\makeatother
\end{thebibliography}




\bsp	
\label{lastpage}
\end{document}